\documentclass{aastex61}
\usepackage{graphicx}

\usepackage{natbib}
\usepackage{mathrsfs}
\usepackage{amsmath}
\usepackage{array}
\newcolumntype{P}[1]{>{\centering\arraybackslash}p{#1}}

\shorttitle{Impulsive-eruptive homologous flares from mini-sigmoid}
\shortauthors{Mitra et al.}

\begin{document}

\title{Eruptive-impulsive homologous M-class flares associated with double-decker flux rope configuration in mini-sigmoid of NOAA 12673}

\correspondingauthor{Prabir K. Mitra}
\email{prabir@prl.res.in}

\author{Prabir K. Mitra}
\affiliation{Udaipur Solar Observatory, Physical Research Laboratory, Udaipur 313 001, India}
\affiliation{Department of Physics, Gujarat University, Ahmedabad 380 009, India}

\author{Bhuwan Joshi}
\affiliation{Udaipur Solar Observatory, Physical Research Laboratory, Udaipur 313 001, India}

\author{Astrid M. Veronig}
\affiliation{Institute of Physics \& Kanzelh\"ohe Observatory, University of Graz, Universit\"{a}tsplatz 5, A-8010 Graz, Austria}

\author{Ramesh Chandra}
\affiliation{Department of Physics, DSB Campus, Kumaun University, Nainital 263 002, India}

\author{K. Dissauer}
\affiliation{Institute of Physics \& Kanzelh\"ohe Observatory, University of Graz, Universit\"{a}tsplatz 5, A-8010 Graz, Austria}
\affiliation{NorthWest Research Associates, 3380 Mitchell Lane, Boulder CO, 80301, USA}

\author{Thomas Wiegelmann}
\affiliation{Max-Planck-Institut f\"ur Sonnensystemforschung, Justus-von-Liebig-Weg 3, D-37077 G\"ottingen, Germany}

\begin{abstract}
We present a multiwavelength analysis of two homologous, short lived, impulsive flares of GOES class M1.4 and M7.3, that occurred from a very localized mini-sigmoid region within the active region NOAA 12673 on 2017 September 7. Both flares were associated with initial jet-like plasma ejection which for a brief amount of time moved toward east in a collimated manner before drastically changing direction toward southwest. Non-linear force-free field extrapolation reveals the presence of a compact double-decker flux rope configuration in the mini-sigmoid region prior to the flares. A set of open field lines originating near the active region which were most likely responsible for the anomalous dynamics of the erupted plasma, gave the earliest indication of an emerging coronal hole near the active region. The horizontal field distribution suggests a rapid decay of the field above the active region, implying high proneness of the flux rope system toward eruption. In view of the low coronal double-decker flux ropes and compact extreme ultra-violet (EUV) brightening beneath the filament along with associated photospheric magnetic field changes, our analysis supports the combination of initial tether-cutting reconnection and subsequent torus instability for driving the eruption.

\end{abstract}


\section{Introduction} \label{sec:intro}

Flares are transient activities occurring in the solar atmosphere in which a huge amount of energy is released within a short time i.e., few minutes to few hours. Earth-directed coronal mass ejections (CMEs) along with their associated eruptive flares are known to produce hazardous effects in the near-Earth environment and drive geomagnetic storms. Magnetic reconnection, a topological reconfiguration of magnetic field in a plasma medium \citep{Priest2000}, is widely accepted to be the fundamental energy release process during solar transient activities. In the process of reconnection, magnetic energy is rapidly converted into plasma heating, bulk motions and kinetic energy of non-thermal particles \citep{Priest2002, Shibata2011}. Eruptive flares and their associated CMEs are, therefore, responsible for large-scale changes in the magnetic structure of the solar atmosphere. With a complex mechanism involving large-scale magnetic fields and its direct consequences on the Earth's atmosphere, flares and CMEs have been widely studied over the years and still is a prominent field of interest \citep[e.g., reviews by][]{Fletcher2011, Benz2017, Green2018}.

The `Standard Flare Model', also known as the CSHKP model \citep{Carmichael1964, Sturrock1966, Hirayama1974, Kopp1976}, considers the existence of a prominence as a pre-requisite for the initiation of eruptive flares. Theoretical models suggests that the basic structure of a prominence/filament is composed of magnetic flux rope (MFR) defined as a set of twisted magnetic field lines wrapped around its central axis more than once \citep{Gibson2006}. Further, the MFR is identified as the dark cavity in the 3-part structure of CMEs \citep[e.g.,][]{Riley2008}. Once the MFR is dynamically activated by an external triggering or by some kind of instabilities and it is set into an eruptive motion, magnetic reconnection begins in a vertical current sheet formed beneath the MFR causing intense flare emission. The CSHKP model successfully incorporates several key features of eruptive flares: flare ribbons; looptop and footpoint sources; hot cusp; post-flare loop arcade; etc. However, the processes that are responsible for the formation of MFR and triggering mechanisms for the eruptive flares are still important and debatable in solar physics. Also, in many eruptive flares, the spatial evolution of looptop sources and flare ribbons during the early phases exhibit significant deviation from the classical scenario described in the CSHKP model which point toward a much complicated energy release process in complex magnetic configuration \citep[e.g.,][see also review by \citealp{Joshi2012}]{Veronig2006, Joshi2009, Dalmasse2015, Joshi2017, Gou2017}.

It is essential to note that the CSHKP model is a 2D model, and therefore, it is not designed to accommodate the 3D structures and configurations e.g., sheared arcades, J-shaped flare ribbons, flux ropes, complex flare loops, etc., which are important in the understanding of solar flares. To implement these features in a general flare model, the CSHKP model has been extended to three dimensions on the basis of a series of numerical simulations \citep{Aulanier2012, Aulanier2013, Janvier2013, Janvier2014}. These MHD simulation results suggest that in the highly sheared preflare magnetic configuration, small-scale current sheets could be generated in the regions of high magnetic gradients i.e., quasi-separatrix layers \citep[QSLs;][]{Titov2002}. Reconnections in these current sheets are responsible for the formation of MFRs from the sheared arcades as well as its destabilization. During the eruption of the MFRs, the inner legs of the sheared arcades which envelop the MFR, straighten vertically beneath the erupting MFR and eventually reconnect resulting in the formation of postflare arcades.

The successful eruption of an MFR is essential for generating a CME. There are two basic groups of models describing the activation and eruption of an MFR from its stable state: models invoking ideal MHD instabilities and models invoking magnetic reconnection \citep[see e.g., reviews by][]{Priest2002, Aulanier2014, Green2018}. In the ideal instability model, a pre-existing MFR can erupt if the background magnetic field displays a rapid decay with height \citep[torus instability;][]{Kliem2006}, or the rope's twist number increases beyond a critical value \citep[kink instability;][]{Torok2004}. Two representative reconnection models, namely, tether-cutting and magnetic breakout, use different preflare magnetic configurations of the active region (AR) while describing the eruption of an MFR. The tether-cutting model requires a bipolar magnetic field configuration, with the earliest reconnection (i.e., triggering process) taking place deep in the sheared core fields \citep{Moore1992, Moore2001}. The breakout model involves a multipolar topology, containing one or more pre-existing coronal null points \citep{Karpen2012}. In this case, the CME onset is triggered by reconnection occurring well above the core region which reduces the tension of the overlying field \citep{Antiochos1999}. Irrespective of the triggering mechanism, once the MFR attains eruptive motion, standard flare reconnection sets in beneath the erupting MFR and this scenario is common to all the models of eruptive flares \citep[e.g., see][]{Vrsnak2004, Veronig2006, Liu2008, Joshi2013, Vrsnak2016, Joshi2016, Veronig2018, Mitra2019, Sahu2020}.

While flares and CMEs result in large-scale changes in the magnetic configuration, coronal jets are relatively small-scale solar eruptive phenomena identified as collimated ejection of plasma in the solar atmosphere \citep[see e.g.,][]{Brueckner1983}. Jets are believed to play an important role in transporting mass and energy from the lower to the upper corona which may have implications in the coronal heating problem \citep[see review by][]{Raouafi2016}. Observations by the Soft X-ray Telescope (SXT) on board Yohkoh \citep{Tsuneta1991} initiated extensive investigations of coronal jets and it was realized that jets are the manifestations of interchange reconnection between a closed and a nearby open magnetic field region \citep[see e.g.,][]{Shibata1992, Shibata1996, Shimojo1996, Bhatnagar1996, Shimojo1998}. Such magnetic configurations can be formed when a magnetic patch emerges in a coronal hole of opposite polarity \citep[``anemone'' type AR; see,][]{Asai2008}. It should be noted that such collimated eruptions which eject into the corona having their bases magnetically rooted in the photosphere \citep{Moore2010}, are observed in different wavelengths and have been named according to the associated observing wavelength regime e.g., EUV jets, H$\alpha$ surges, EUV and H$\alpha$ macrospicules etc. \citep[see e.g.,][]{Moore1977, Schmieder1995, Jiang2007}. \citet{Moore2010} proposed a dichotomy in solar jets: standard and blowout. In the standard jet scenario, reconnection between a pre-existing open field and a newly emerging magnetic field of opposite polarity, is responsible for guiding the hot plasma along the post reconnection open field resulting in a narrow, long jet-spire. On the other hand, the blowout category of jets involve eruption of the jet's base-arch that contains a mini- flux rope, resulting in a broader and apparently untwisting jet-spire and a CME \citep[see also,][]{Archontis2013, Joshi2016, Chandra2017b, Joshi2017b}.

During 2017, when the Sun was moving toward the minimum phase of the solar cycle 24, a simple $\alpha$-type AR NOAA 12673 emerged on the eastern limb of the Sun on 2017 August 28. It gradually became complex with time and turned into a $\beta\gamma\delta$-type sunspot on 2017 September 5. Before disappearing over the western limb of the Sun on 2017 September 10, it produced 4 X-class and 27 M-class flares along with numerous C-class flares making it one of the most powerful ARs of solar cycle 24. Notably, it produced the two biggest flares of solar cycle 24, namely the X9.3 event on 2017 September 6 and the X8.2 flare on 2017 September 10, which were subject to numerous studies \citep[e.g.,][]{Yang2017, Seaton2018, Romano2018, Verma2018, Guo2018, LiuW2018, Hou2018, Gary2018, LiuL2018b, Veronig2018, Mitra2018, Romano2019, Moraitis2019, Liu2019, Duan2019, Chen2020}. Most of the flaring activity from the AR occurred from the central region where the sunspot group arranged into `$\delta$'-configuration \citep{Kunzel1960}. This region was characterized by a sharp polarity inversion line (PIL) exhibiting high magnetic gradient across the PIL; for example, \citet{Mitra2018} noted a magnetic gradient of 2.4 kG Mm$^{-1}$ in the line of sight (LOS) magnetic field across the PIL, prior to the X-flares flares of 2017 September 6.

In this article, we present a comprehensive multiwavelength analysis of two impulsive flares of classes M1.4 and M7.3, which occurred on 2017 September 7 in a very localized region situated near the edge of the main sunspot group of AR 12673. Both events were accompanied with highly collimated eruptions, a characteristic of coronal jet. The jet-flare events initiated from an unusually small coronal sigmoid which we explore in detail by (E)UV imaging and coronal magnetic field modeling; thanks to the high resolution data from the \textit{Solar Dynamics Observatory} \citep[\textit{SDO};][]{Pesnell2012}. An important aspect of this study lies in the early dynamics of the eruptions during the flares. Section \ref{data} provides a brief account of the observational data and analysis techniques. In Section \ref{results}, we derive the observational results on the basis of measurements taken at photospheric, chromospheric and coronal levels. In Section \ref{extpl}, we compare the chromospheric and coronal observations of different flare-associated features with the modeled coronal magnetic configurations. We discuss and interpret our results in Section \ref{discussion}.

\section{Observational data and analysis techniques} \label{data}
For imaging the solar atmosphere, we used observations from the Atmospheric Imaging Assembly \citep[AIA;][]{Lemen2012} on board \textit{SDO}. AIA produces 4096$\times$4096 pixel full disk solar images with a spatial resolution of $1\farcs5$ and pixel scale of $0\farcs6$ in 10 (E)UV channels originating in different heights of the solar atmosphere. Particularly, the 12 s cadence observations in the 94 \AA\ (Fe XVIII; $log(T)=6.8$) and 335 \AA\ (Fe XVI; $log(T)=6.4$) filters along with the 193 \AA\ (Fe XII, XXIV; $log(T)=6.2, 7.3$) channel were used for investigation of coronal activities associated with the flares. For imaging of the lower atmospheric layers, we have extensively used 12 s cadence observations in the 304 \AA\ (He II; $log(T)=4.7$) and 24 s cadence observations in the 1600 \AA\ (C IV \& continuum; $log(T)=5 ~\& ~3.4$) channels.

\textit{Reuven Ramaty High Energy Solar Spectroscopic Imager}
\citep[\textit{RHESSI};][]{Lin2002} observed the second (M7.3) flare almost completely and the interval between the two homologous flares while it missed the first (M1.4) event due to the spacecraft's passage through the South Atlantic Anomaly (SAA\footnote{\url{https://heasarc.gsfc.nasa.gov/docs/rosat/gallery/misc_saad.html}}). \textit{RHESSI} observed the full Sun with an unprecedented combination of angular (spatial) resolution (as fine as $\approx2\farcs3$) and energy resolution (1--5 keV) in the energy range 3 keV--17 MeV. The image reconstruction is done with the CLEAN algorithm \citep{Hurford2002} using only front detector segments with an integration time of 12 s and pixel scale of $2\farcs0$. Out of 9 detector segments, segment 2 was excluded for imaging at 6--12, 12--25, 25--50, and 50--100 keV, while segments 2 and 7 were excluded for 3--6 keV imaging.

Photospheric structures associated with the AR NOAA 12673 were observed from full disk 4096$\times$4096 pixel intensity images and line-of-sight (LOS) magnetograms from the Helioseismic and Magnetic Imager \citep[HMI;][]{Schou2012} on board \textit{SDO} at a spatial resolution of $1\farcs0$ and pixel scale of $0\farcs5$ and 45 s cadence. To investigate the coronal magnetic structures associated with the AR, we used a global potential field source surface model \citep[PFSS\footnote{\url{https://www.lmsal.com/~derosa/pfsspack/}};][]{Wang1992} and a non-linear force-free field (NLFFF) model \citep{Wiegelmann2010, Wiegelmann2012}. As boundary condition for the NLFFF modeling, we used photospheric vector magnetograms of 2017 September 07 09:46 UT from the `hmi.sharp\_cea\_720s' series of \textit{SDO}/HMI at a reduced resolution of $0\fdg06$ and a temporal cadence of 720 s. The NLFFF extrapolations were done in a Cartesian volume of dimensions 344$\times$224$\times$224 pixels which corresponds to a physical size of $\approx$250$\times$160$\times$160 Mm$^3$ in the solar atmosphere. The NLFFF field lines were visualized using the Visualization and Analysis Platform for Ocean, Atmosphere, and Solar Researchers \citep[VAPOR\footnote{\url{https://www.vapor.ucar.edu/}};][]{Clyne2007} software which produces an interactive 3D visualization environment.

We calculated the magnetic decay index in the NLFFF extrapolation volume using the results of  potential field extrapolation obtained by solving a Green's function method \citep{Seehafer1978}. The magnetic decay index ($n$) is given by the equation: $n=-\frac{log(B_{ex}(z))}{log(z)}$, where $B_{ex}(z)$ is the horizontal component of the external field above the AR and $z$ is height \citep{Bateman1978, Kliem2006}.
 
The CME that originated from the filament eruption was observed by the C2 and C3 instruments of the Large Angle and Spectrometric Coronagraph \citep[LASCO;][]{Brueckner1995} on board the \textit{Solar and Heliospheric Observatory} \citep[\textit{SOHO};][]{Domingo1995}. C2 and C3 are white-light coronagraphs that image the solar corona from 1.5--6 $R_\odot$ and from 3.7--30 $R_\odot$, respectively.

\section{Mutiwavelengths observations and analysis} \label{results}
\subsection{Event overview} \label{evov}

\begin{figure}
   \centering
   \epsscale{1.2}
   \plotone{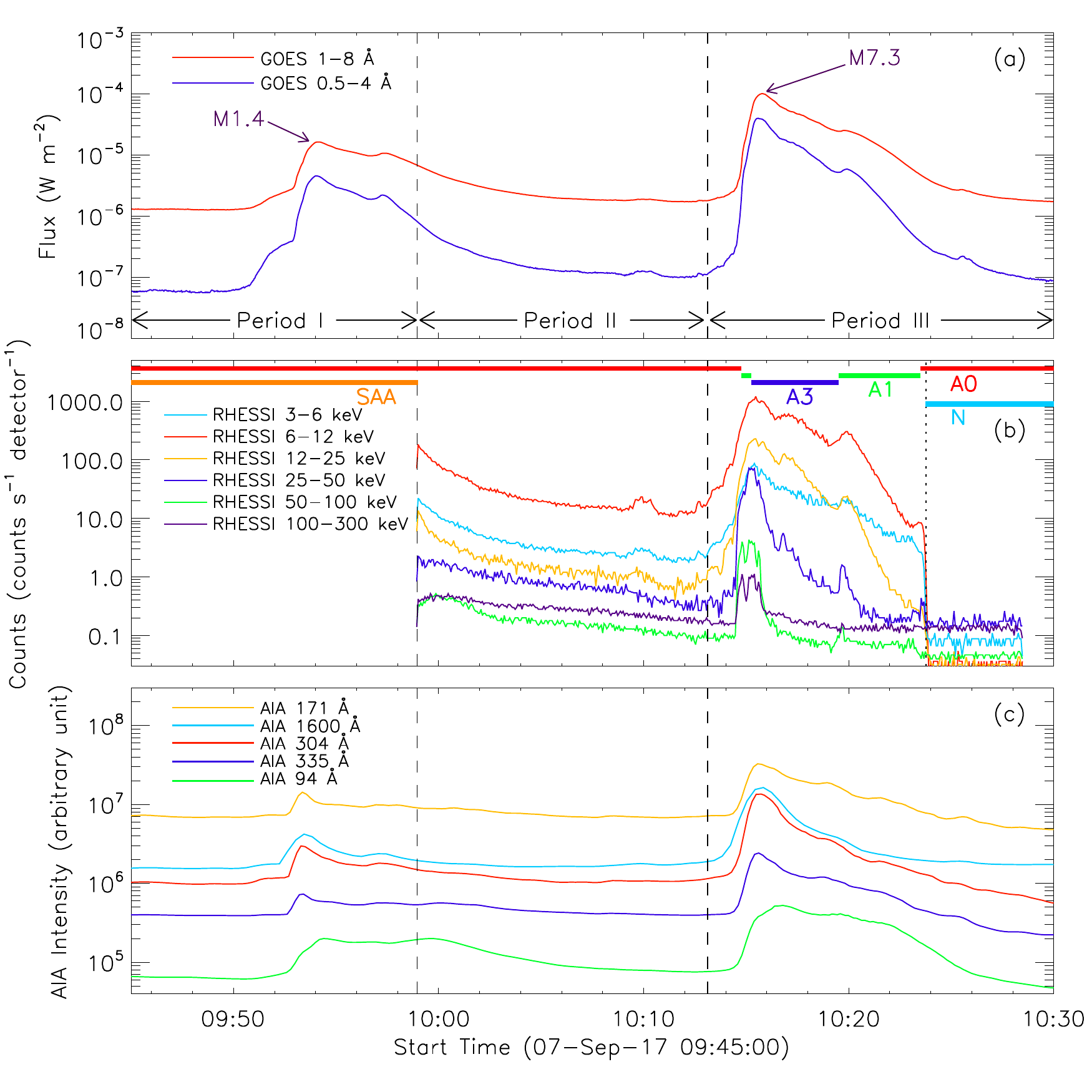}
   \vspace{-0.9 cm}
   \caption{Panel (a): GOES SXR flux in the of 1--8 \AA\ (red curve) and 0.5--4 \AA\ (blue curve) bands on 2017 September 7 from 09:45 UT to 10:30 UT covering the two M-class flares under study. Panel (b): \textit{RHESSI} X-ray fluxes normalized by factors of $\frac{1}{10}$, $\frac{1}{30}$, $\frac{1}{100}$, and $\frac{1}{100}$ for 12--25 keV, 25--50 keV, 50--100 keV, and 100--300 keV energy bands, respectively. The horizontal red, green, and blue bars in panel (b) indicate the \textit{RHESSI} attenuator states (A0, A1, and A3, respectively). The orange and blue horizontal bars in panel (b) indicate the durations missed by \textit{RHESSI} due to SAA and `\textit{RHESSI}-night', respectively. Panel (c): AIA (E)UV lightcurves. For clear visualization, AIA lightcurves in the channels 171 \AA\ and 94 \AA\ are scaled by $\frac{1}{2}$ and $\frac{1}{10}$, respectively.}
   \label{goes}
\end{figure}

We present observations of AR NOAA 12673 on 2017 September 7 from 09:45 UT to 10:30 UT. During this period, the AR produced two GOES M-class flares\footnote{\url{http://www.lmsal.com/solarsoft/ssw/last_events-2017/last_events_20170908_1158/index.html}}. The temporal evolution of the M-class flares is represented by the soft X-ray (SXR) flux variation in the 1--8 \AA\ channel of GOES  in Figure \ref{goes}(a) (shown by the red curve). The first M-class flare (M1.4; indicated by an arrow in Figure \ref{goes}(a)) initiated at $\approx$09:51 UT. After a brief period of slow rise from $\approx$09:51 UT to $\approx$09:53 UT, which is often observed to precede the impulsive flare phase \citep[see e.g.,][]{Veronig2002, Mitra2019}, the flux in both GOES SXR channels (i.e., 1--8 \AA\ and 0.5--4 \AA ) experienced a rapid enhancement till the peak of the flare at $\approx$09:54 UT. The decay phase of the M1.4 flare is characterized by a subtle rise at $\approx$09:57 UT in an otherwise steady decay of both GOES SXR fluxes. The M7.3 flare was characterized by a brief impulsive phase when the flux in either GOES channels experienced an abrupt rise at $\approx$10:14 UT. The flare peaked at $\approx$10:16 UT and thereafter decayed until the end of our observing period at 10:30 UT when the SXR fluxes attained the level of the corresponding preflare backgrounds. \textit{RHESSI} observed the AR from 2017 September 7 $\approx$09:59 UT to $\approx$10:24 UT which almost fully covered the M7.3 flare and the interval between the two flares (Figure \ref{goes}(b)). Depending on the time evolution of the GOES SXR fluxes combined with the available \textit{RHESSI} observations, we divided the entire duration (09:45 UT--10:30 UT) into three periods: periods I and III cover the M1.4 flare and the M7.3 flares with the associated eruptions, respectively, whereas period II covers the rather quiet phase in between the two M-class flares (Figure \ref{goes}(a)).

The AIA (E)UV lightcurves from the AR during this time (Figure \ref{goes}(c)) show a general agreement with the GOES SXR flux variation signifying that the disk-integrated GOES measurement was largely influenced by the coronal activity of the AR NOAA 12673. The intensity variations in all the AIA channels suggest the initiation of the M1.4 flare at $\approx$09:52 UT and the M7.3 flare at $\approx$10:14 UT. We note similar intensity variations in the light curves of different AIA channels. We have summarized the different phases of the flares in the studied interval in Table \ref{table}.

\begin{figure}
   \centering
   \epsscale{0.9}
   \plotone{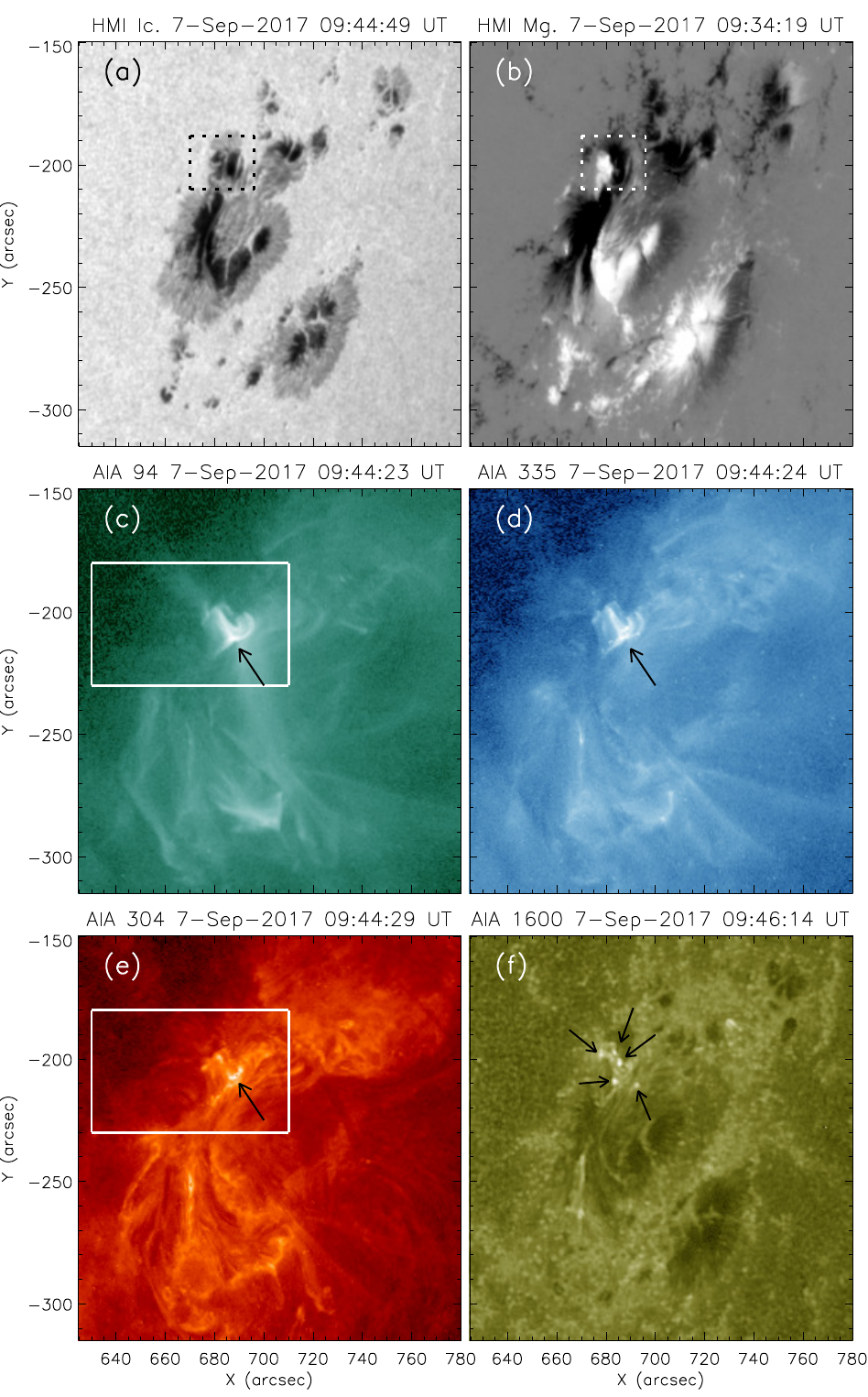}
   \caption{Panel (a): HMI white light image of AR 12673 on 2017 September 7 09:44 UT. Panel (b): Co-temporal HMI magnetogram. Panels (c)--(e): AIA EUV images of the AR in 94 \AA , 335 \AA , and 304 \AA\ respectively, showing the morphology of the active region in the corona and chromosphere. Panel (f): AIA UV image of the active region in 1600 \AA\ . The dashed boxes in panels (a) and (b) indicate the photospheric region associated with the two M-class flares. The boxes in panels (c) and (e) indicate the field of view of the AR plotted in Figures \ref{m1_4}, \ref{m7_3_2}, and \ref{m7_3}. Arrows in panels (c)--(e) indicate the mini-sigmoid region whereas the arrows in panel (f) indicate brightenings at the location of the mini-sigmoid.}
   \label{intro}
\end{figure}

Figure \ref{intro} displays the morphology and configuration of the AR at the photosphere and different layers above it prior to the reported events. Comparison between the continuum image and magnetogram of the AR (Figures \ref{intro}(a) and (b)) suggests that the southwestern sunspot group was primarily of positive polarity while negative polarity dominated the northern dispersed sunspot group. Our interest lies in the central part of the AR which displayed a complex, bipolar configuration. The coronal images of the AR show a very interesting feature in the northern part of the central sunspot group (see the region marked by dashed box in Figures \ref{intro}(a) and (b)), with a shape of a semicircular arc which is indicated by the black arrow in Figures \ref{intro}(c)--(e). Careful observation suggests the shape of this bright feature to be similar to an inverted `S'. Comparison between coronal and photospheric images of the AR unveils that the structure was lying over the bipolar sunspot region (cf. the region indicated in the dashed box in Figures \ref{intro}(a) and (b)). This inverted `S' shaped bright structure can be thought of as a mini-sigmoidal region. In AIA UV images (Figure \ref{intro}(f)), though, the mini-sigmoid is not clearly visible, few very localized bright dots can be found at the same location (indicated by the black arrows in Figure \ref{intro}(f)). In the next three Sections (i.e., Section \ref{fm1_4}, \ref{fp2}, and \ref{fm7_3}), we focus on the region shown in the boxes in Figures \ref{intro}(c) and (e) as both the M-class flares reported in this article occurred in this region.

\begin{deluxetable*}{P{0.5cm}P{4cm}P{1.4cm}P{3.0cm}P{7.9cm}}
\tablenum{1}
\tablecaption{Chronology of events during the two M-class flares occurred on 2017 September 7}
\label{table}
\tablehead{
\colhead{Sr.} & \colhead{Evolutionary} & \colhead{} & \colhead{Observing} & \colhead{}\\
\colhead{No.} & \colhead{stages} & \colhead{Time} & \colhead{instrument/ wavelength} & \colhead{Remarks}
}

\startdata
1 & Preflare phase & 09:45 UT & AIA (E)UV \& HMI & A very localized inverted `S' structured brightening was observed in AIA 304 and other AIA coronal images in the AR. We call it ``mini-sigmoid''. \\
2 & Initiation of the M1.4 flare & 09:49 UT & GOES 1--8 \AA\ & \\
3 & Initiation of plasma ejection during the M1.4 flare & 09:53 UT & AIA 94 \AA\ and 304 \AA\ & Collimated ejection of plasma from the core of the mini-sigmoid toward east. \\
4 & End of the M1.4 flare & 09:58 UT & GOES 1--8 \AA\ & A small filament appeared in the mini-sigmoid region. \textit{RHESSI} observation started at $\approx$09:59 UT. \\
5 & Initiation of the M7.3 flare & 10:11 UT & GOES 1--8 \AA\ & \\
6 & Initiation of the first phase of plasma ejection during the M7.3 flare & 10:14 UT & AIA 304 \AA\ & Collimated plasma was ejected from the core of the mini-sigmoid toward east. \textit{RHESSI} sources of energies up to $\approx$100 keV were found from the mini-sigmoid region. \\
7 & Initiation of the second phase of plasma ejection during the M7.3 flare & 10:16 UT & AIA 304 \AA\ & The southern end of the filament erupted ejecting plasma toward south. \\
8 & Merging of the plasma ejected in the first phase with plasma ejected in the second phase & 10:22 UT & AIA 335 \AA\ & Plasma ejected toward east during the M7.3 flare strangely changed direction from east to south-west and merged with the plasma ejected in the second phase during the M7.3 flare. \\
9 & End of the M7.3 flare & 10:18 UT & GOES 1--8 \AA\ & \\
10 & Detection of CME & 10:24 UT & LASCO C2 & CME propagated along the central PA of 254$^\circ$ and angular width of 32$^\circ$ with a linear speed of 470 km s$^{-1}$.\\
\enddata
\end{deluxetable*}

\subsection{Period I: Evolution of the M1.4 flare} \label{fm1_4}
\begin{figure}
   \centering
   \epsscale{1.0}
   \plotone{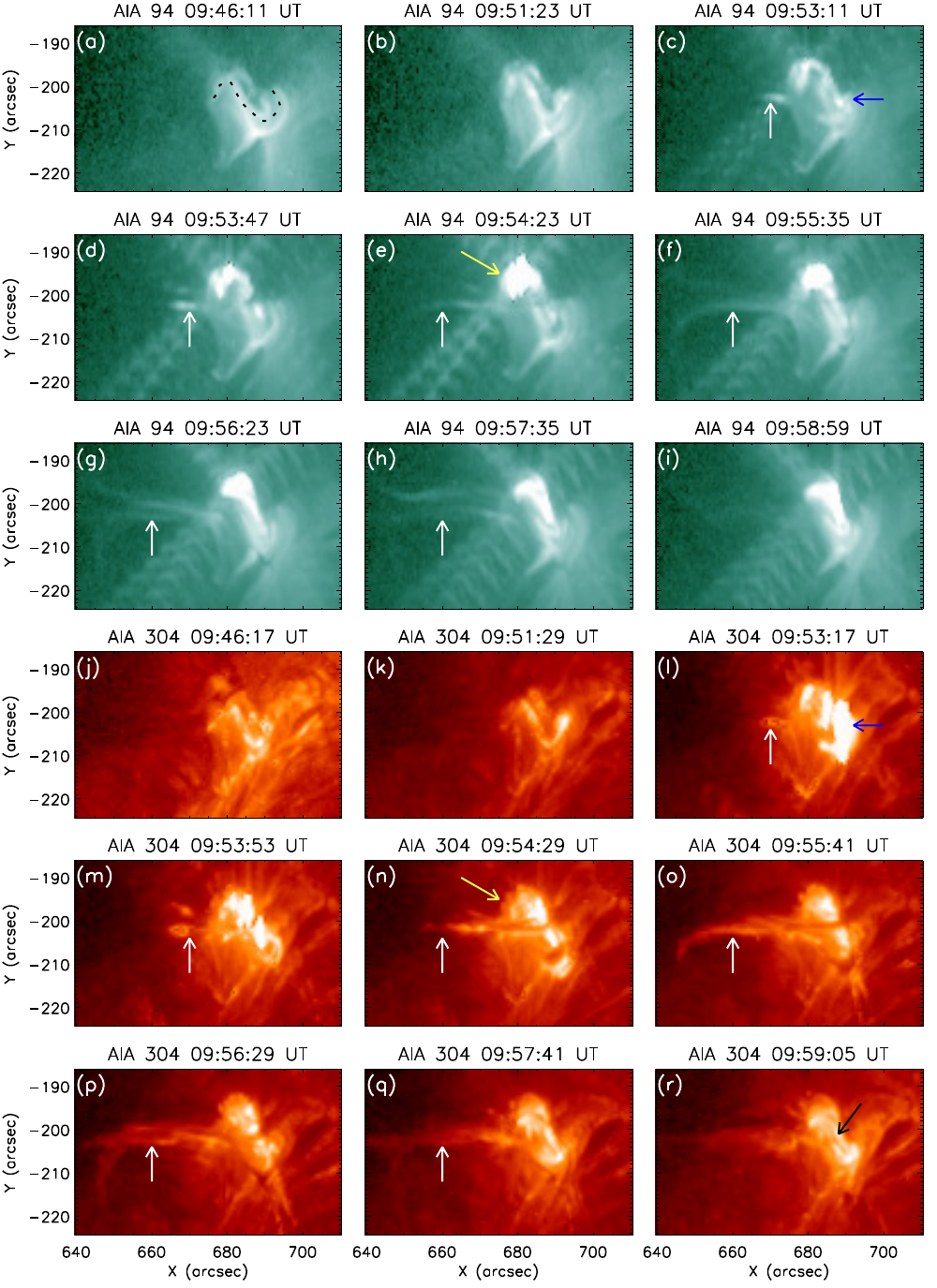}
   \caption{Series of AIA 94 \AA\ (panels (a)--(i)) and 304 \AA\ (panels (j)--(r)) images showing the time evolution of the M1.4 flare. The mini-sigmoid structure identified in the preflare phase is outlined by the dashed black dotted curve in panel (a). The white arrows in different panels indicate the ejected jet-like plasma. The blue arrows in panels (c) and (l) indicate a newly emerged brightening in the western end of the mini-sigmoid which led to the flare onset. The yellow arrows in panels (e) and (n) indicate the brightening in the eastern end of the mini-sigmoid during the peak phase of the M1.4 flare. The black arrow in panel (r) indicate a newly formed filament structure during the decaying phase of the M1.4 flare.}
   \label{m1_4}
\end{figure}

Figure \ref{m1_4} displays a series of AIA 94 \AA\ (Figures \ref{m1_4}(a)--(i)) and AIA 304 \AA\ (Figures \ref{m1_4}(j)--(r)) images showing the morphological evolution of the region shown inside the boxes in Figures \ref{intro}(c) and (e) during the M1.4 flare and the associated jet-like plasma eruption. As discussed in Section \ref{evov}, before the flare onset, the northern part of the AR contained a mini-sigmoid. In terms of sharpness of the observed feature and its relative brightness in comparison to the ambient medium, the mini-sigmoid was apparently more prominent in the hot, coronal AIA 94 \AA\ channel than
the relatively cooler AIA 304 \AA\ filter. In Figure \ref{m1_4}(a), we outline the mini-sigmoid by a black dashed line which brightened up after 09:50 UT, marking the onset of the M1.4 flare. After 09:53 UT, we observe a localized kernel-like brightening in the western leg of the mini-sigmoid (shown by the blue arrow in Figure \ref{m1_4}(c)). Further, looking at the spatial extent as well as the relative intensity, this localized brightening was observed to be more prominent in the 304 \AA\ than in the 94 \AA\ observations (cf. Figures \ref{m1_4}(c) and (l)) which suggests that the early energy release during the initiation of M1.4 flare occurred in lower i.e., chromospheric heights (cf. Figures \ref{m1_4}(c) and (l)). At the same time we observe the ejection of plasma from the middle of the mini-sigmoid (shown by the white arrow in Figure \ref{m1_4}(c)). The ejected plasma followed a very narrow and collimated path (i.e., a jet-like eruption) toward the east for a distance of $\approx$40 arcsec and then abruptly changed its direction. The progress of the ejected plasma is indicated by the white arrows in Figures \ref{m1_4}(c)--(h) and (l)--(q). Here, we note that the ejecting plasma was observed more clearly in AIA 304 \AA\ images than in AIA 94 \AA\ images. This is indicative of cooler plasma being ejected from lower layers in the solar atmosphere (presumably filament material), and partially being heated during the ejection process. At the peak phase of the flare, the eastern leg of the mini-sigmoid became the brightest location in the entire AR as observed in the AIA 94 \AA\ images (Figure \ref{m1_4}(e); shown by the yellow arrow). The decay phase of the M1.4 flare was characterized by the appearance of a small filament structure observed in AIA 304 \AA\ images (shown by the black arrow in Figure \ref{m1_4}(r)).

\subsection{Period II: Quiet phase between the two M-class flares} \label{fp2}
\begin{figure}
   \centering
   \epsscale{1.0}
   \plotone{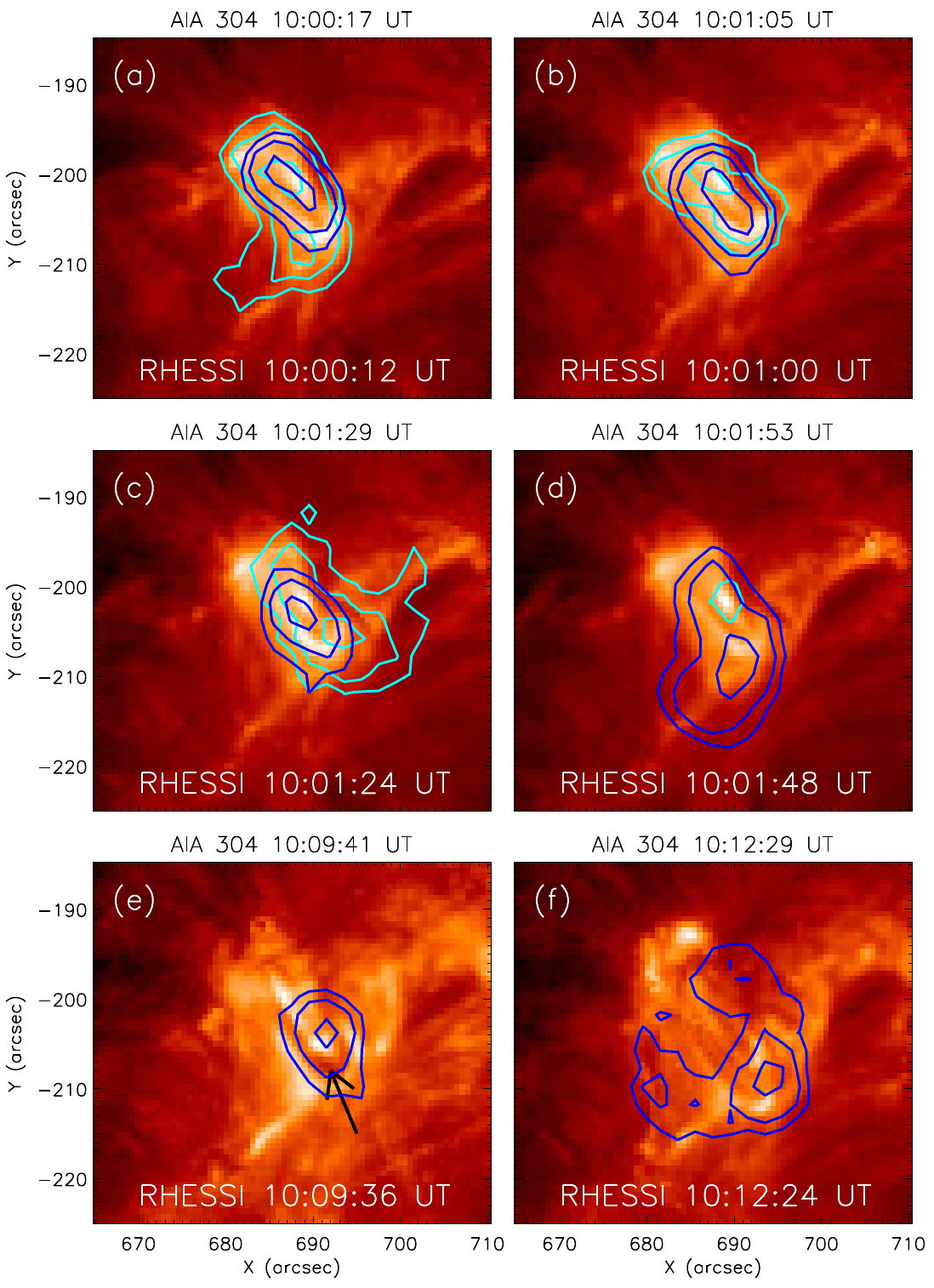}
   \caption{Series of AIA 304 \AA\ images of the flaring region during the relatively quiet period between the two M-class flares. Contours of \textit{RHESSI} 3--6 keV (sky) and 6--12 keV (blue) are overplotted on each panel. Contour levels are set as 70\%, 80\%, and 95\% of the corresponding peak flux. The arrow in panel (e) indicate a developing filament from the mini-sigmoid structure prior to the onset of the M7.3 flare.}
   \label{p2}
\end{figure}

A small filament started to appear during the late phase of the M1.4 flare along the axis of the mini-sigmoid (Section \ref{fm1_4}). After the M1.4 flare, i.e., in `Period II' of the study (see Figure \ref{goes}), this filament became prominent. Notably, after missing Period I due to the SAA, \textit{RHESSI} started observing in period II, with high sensitivity at low energies (no attenuator in place, state A0). In Figure \ref{p2}, we show a series of AIA 304 \AA\ images overplotted with contours of \textit{RHESSI} X-ray sources in the energy ranges 3--6 keV (shown by sky contours) and 6--12 keV (shown by blue contours). We readily find emission in the 6--12 keV range to have peak intensity along the axis of the mini-sigmoid (Figure \ref{p2}(a)). We also note that the X-ray emission in the 3--6 keV energy range at 10:00 UT displayed two distinct sources on either sides of the filament. At $\approx$10:10 UT (Figure \ref{p2}(e)), when the filament was very prominently visible (shown by the black arrow), we find very localized X-ray emission in the 6--12 keV range. At 10:12 UT, the 6--12 keV emission was characterized by an intense source at the top of the filament and two weaker sources on the either sides of the filament (Figure \ref{p2}(f)).

\subsection{Period III: Evolution of the M7.3 flare and associated filament eruption} \label{fm7_3}

\begin{figure}
   \centering
   \epsscale{1.0}
   \plotone{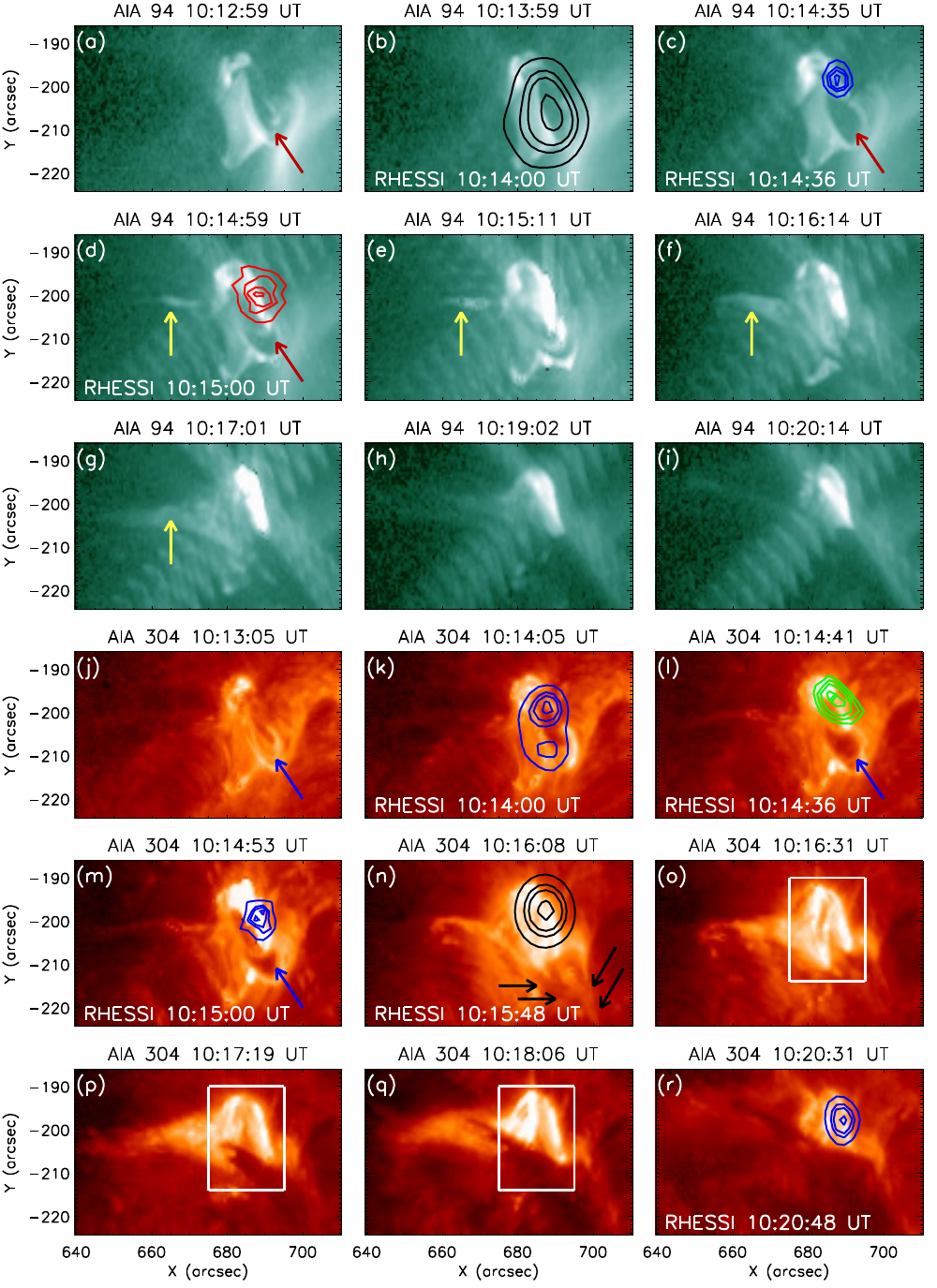}
   \caption{Series of AIA 94 \AA\ (panels (a)--(i)) and 304 \AA\ (panels (j)--(r)) images showing the time evolution of the M7.3 flare. The brown and blue arrows in different panels indicate the development of the filament. The yellow arrows in panels (d)--(g) indicate the first phase of plasma ejection during the M7.3 flare. The black arrows in panel (n) indicate a second phase of the eruption during the flare. The white boxes in panels (o), (p) and (q) indicate post flare arcade observed from an edge-on view. Contours of \textit{RHESSI} 6--12 keV (blue), 12--25 keV (black), 25--50 keV (red), and 50--100 keV (green) are overplotted in different panels. Contour levels are set as 50\%, 70\%, 80\%, and 95\% of the corresponding peak flux for 6--12 keV, 12--25 keV, and 25--50 keV energy bands and 60\%, 70\%, 80\%, and 95\% of the corresponding peak flux for the 50--100 keV band.}
   \label{m7_3_2}
\end{figure}

\begin{figure}
   \centering
   \epsscale{1.}
   \plotone{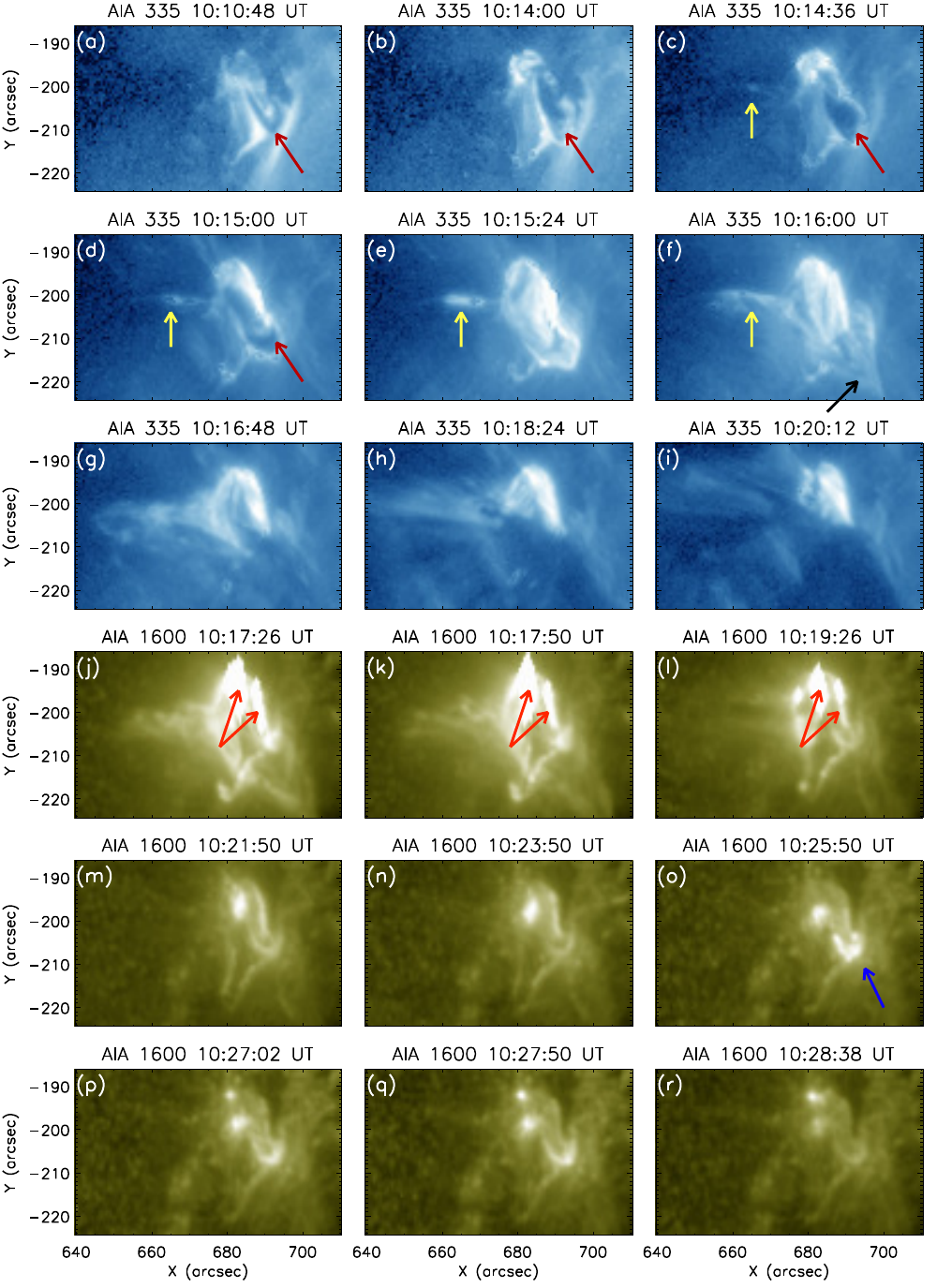}
   \caption{Series of AIA 335 \AA\ (panels (a)--(i)) and 1600 \AA\ (panels (j)--(r)) images showing the time evolution of the M7.3 flare. Brown arrows in panels (a)--(d) indicate emergence of a small filament structure in the preflare phase. Yellow arrows in panels (c)--(e) indicate the ejecting plasma. Black arrow in panel (f) indicate the direction a second stream of ejecting plasma. The red arrows in panels (j)--(l) indicate flare ribbon like structures formed during the M7.3 flare. The blue arrow in panel (o) indicate a subtle brightening occurred in the decay phase of the flare which, most probably, was responsible for the small enhancement in the GOES SXR lightcurves (see Figure \ref{goes}).}
   \label{m7_3}
\end{figure}

The M7.3 flare was initiated at $\approx$10:14 UT (cf., Figure \ref{goes}) when the newly formed filament was distinctly visible in the mini-sigmoid region both in coronal and chromospheric images (see Figures \ref{m7_3_2} and \ref{m7_3} where the filament is indicated by brown and blue arrows, respectively). The flare brightening was first observed at the northern end of the filament which was followed by a jet-like plasma ejection at $\approx$10:15 UT (indicated by the yellow arrows in Figure \ref{m7_3_2}(d)--(g)). The jet-like plasma ejection was observed slightly earlier in the AIA 335 \AA\ channel than in the AIA 94 \AA\ channel (cf. Figures \ref{m7_3_2}(c) and \ref{m7_3}(c)). However, the eruption was visible in the chromospheric AIA 304 \AA\ images $\approx$1 min earlier than in the AIA 94 \AA\ images sampling hot coronal plasma at $\sim$6 MK (cf. Figure \ref{goes}(c), \ref{m7_3_2}(k), (l) and \ref{m7_3_2}(d)). It is noteworthy that the motion and direction of the initial phase of plasma ejection was very similar to that during the earlier M1.4 flare. The filament brightened up after $\approx$10:15 UT causing the peak phase of the M7.3 flare. Hard X-ray emission up to $\approx$100 keV was observed from the mini-sigmoid region by \textit{RHESSI} during this flare. In Figure \ref{m7_3_2}, we plot contours of \textit{RHESSI} X-ray emission in different energy channels and find X-ray sources to be co-spatial with the EUV brightenings. AIA 304 \AA\ images of the region displayed an interesting feature during this time. The southern part of the filament kept rising and slowly developed into a cusp-like structure at $\approx$10:16 UT (the cusp-like structure is indicated by the black arrows in Figure \ref{m7_3_2}(n)). The structure was also visible in the co-temporal AIA 335 \AA\ channel images (Figure \ref{m7_3}(f); indicated by the black arrow). A second phase of plasma eruption was initiated at $\approx$10:16 UT from the cusp-like structure. Simultaneous eruption of collimated plasma during the two phases partially occulted the bright flare emission from the AR thereafter. Observation of the sigmoidal region after 10:16 UT was only possible in those AIA channels which image the lower atmosphere of the Sun (i.e., 1600 \AA ). We observed two bright ribbon-like structures on either sides of the filament in AIA 1600 \AA\ images until 10:20 UT (indicated by the red arrows in Figures \ref{m7_3}(j)--(l)). However, co-temporal AIA 304 \AA\ images of the same location suggests that the bright structures were associated with a post-flare arcade viewed from an edge-on angle (outlined by the white boxes in Figures \ref{m7_3_2}(o)--(q)). Therefore, we conclude that the bright structures observed in the AIA 1600 \AA\ images in Figures \ref{m7_3}(j)--(l) were a mixture of emissions coming from both the flare ribbons and post-flare arcades. After this time, from the AIA 1600 \AA\ images, no significant morphological changes were observed in the AR till the end of our observing period except a short lived brightening found in the western end of the sigmoid region at $\approx$10:26 UT (shown by the blue arrow in Figure \ref{m7_3}(o)). Here, we remember that a subtle peak was observed in the decaying SXR fluxes of both GOES channels at the same time (see Figure \ref{goes}).

\subsection{Development of the CME from the erupting filament} \label{sec:cme}

\begin{figure}
   \centering
   \epsscale{1.2}
   \plotone{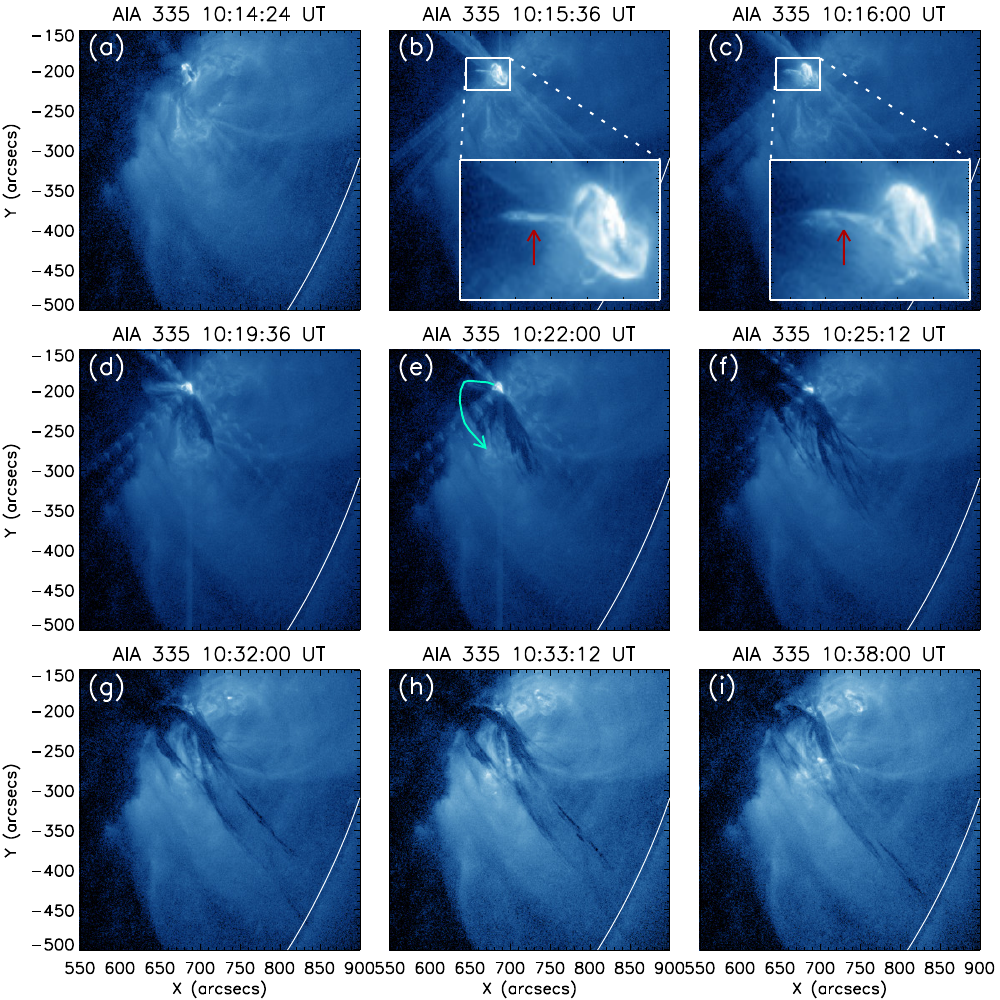}
   \caption{Series of AIA 335 \AA\ images showing large scale eruption of plasma from active region NOAA 12673 during the M7.3 flare reported in this paper. The red arrows in panels (b) and (c) indicate the initial phases of the ejecting plasma moving toward east. The curve in panel (e) outlines the unusual turning of the ejecting plasma from east to south-west. The ejected plasma during the M7.3 flare resulted in a CME of medium speed and small angular width.}
   \label{erup}
\end{figure}

As discussed in Sections \ref{fm1_4} and \ref{fm7_3}, both the M-class flares were associated with plasma ejections. Interestingly, plasma ejection signatures during the M1.4 flare became too weak to be observed few minutes after its first appearance within AIA field of view (FOV). Erupting plasma during the M7.3 flare, however, was distinctly observed to produce a CME by SOHO/LASCO. In this Section, we focus on the motion of the ejected plasma during the M7.3 flare (Figure \ref{erup}) and the corresponding CME (Figure \ref{cme}). The plasma ejection was initiated in a collimated manner toward east (shown by the red arrow in Figures \ref{erup}(b) and (c)) and then dramatically changed its direction toward south-west. In Figure \ref{erup}(e), we have approximately outlined the changing direction of erupting plasma. A second phase of plasma ejection initiated from the western end of the mini-sigmoid region after $\approx$10:16 UT (see Section \ref{fm7_3}). Plasma that was ejected in both of these two phases (Figures \ref{erup}(d)--(i)) during the M7.3 flare, thereafter proceeded together to constitute a CME.

Figure \ref{cme} displays a series of running difference images by the LASCO C2 (Figure \ref{cme}(a)--(b)) and C3 (Figure \ref{cme}(c)) coronagraphs, where the CME that developed from the plasma ejection during the M-class flares can be observed. According to the LASCO CME catalogue\footnote{\url{https://cdaw.gsfc.nasa.gov/CME_list/UNIVERSAL/2017_09/yht/20170907.102406.w032n.v0470.p244g.yht}} \citep{Yashiro2004}, C2 detected the CME at 10:24 UT at $\approx$2.4 R$_\odot$ and it was observed in the field of view of C3 until 16:18 UT when the leading edge of the CME reached $\approx$17.0 R$_\odot$. The narrow CME (angular width being only 32$^\circ$) propagated along the position angle 254$^\circ$ with a linear speed of $\approx$470 $km~ s^{-1}$.

\subsection{Structure and evolution of the magnetic configuration of AR 12673} \label{fl_cncl}
The distribution and configuration of the photospheric magnetic flux of AR 12673 (Figure \ref{hmi_lightcurve}(a)) remained without any major changes during our observing period. However, it experienced consistent changes in the magnetic field strength. In Figure \ref{hmi_lightcurve}(b), we plot the photospheric LOS magnetic flux variation associated with the flaring region shown by the dashed box in Figure \ref{hmi_lightcurve}(a) on 2017 September 07 from 08:00 UT to 10:30 UT. Notably, this region was associated with the formation and eruption of the filament in the mini-sigmoid. We find that, the negative flux underwent a monotonic decrease during the preflare period from $\approx$08:40 UT to $\approx$09:49 UT with a decay rate of $\approx$2.41$\times$10$^{16}$ Mx s$^{-1}$. On the other hand, the positive flux underwent a gradual enhancement during the preflare period with two distinguishable phases of flux decrease ($\approx$08:35 UT--08:40 UT and $\approx$09:17 UT--09:28 UT). The decrease in magnetic flux can be interpreted as observational signature of the photospheric flux cancellation. Flux variation during the flaring intervals (indicated by the dashed and dashed-dotted lines) are quite drastic which seems to be driven by the flaring activity. In this context, it is noteworthy that the second flare was a white light event (WLF) which may produce artefacts in the magnetic field measurements\footnote{During the impulsive phases of WLFs, sudden changes in the LOS photospheric fields have been observed by several earlier studies \citep[see e.g.,][]{Zhao2009, Maurya2009, Maurya2012, Kushwaha2014}. The sudden transient changes observed in LOS magnetograms can be interpreted in terms of the theoretical calculations of \citet{Ding2002} that show the field reversal during strong flares could be an “observational artifact” that is locally induced by bombardment of energetic electron beams at the photosphere.}. 

\section{Magnetic field modeling} \label{extpl}

\subsection{Large-scale magnetic field configuration} \label{pfss}

\begin{figure}
   \centering
   \epsscale{0.45}
   \plotone{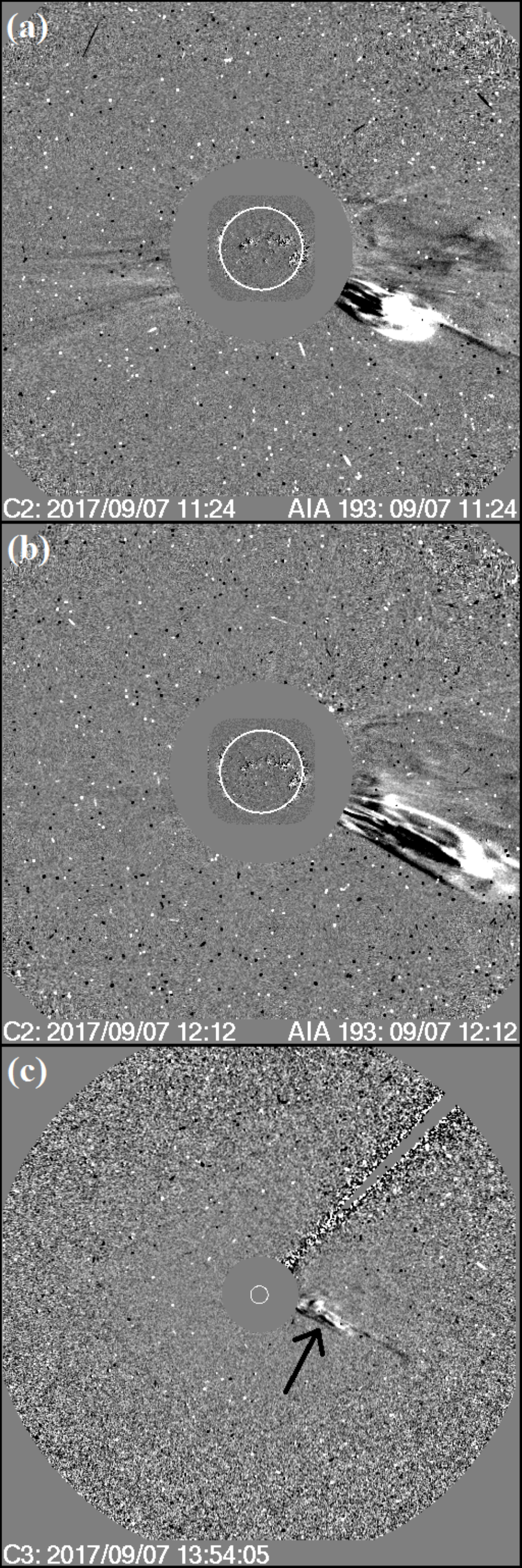}
   \caption{LASCO observations of the CME that developed from the plasma ejection during the M7.3 flare. Panels (a)--(b) present observations from C2 coronagraph (1.5--6 $R_\odot$) while panel (c) shows observation from C3 coronagraph (3.7--30 $R_\odot$). The CME is indicated by the black arrow in panel (c). The CME was first observed by LASCO at 10:24 UT and was observed until 16:18 UT.}
   \label{cme}
\end{figure}

Large-scale magnetic field configurations, such as open magnetic field lines represented by coronal holes \citep[CHs;][]{Cranmer2009}, may strongly influence the early propagation of CMEs, and may cause significant deflections of their original direction of motion \citep{Gopalswamy2009, Heinemann2019}. In order to check if the direction of the CME was influenced by the open field configuration associated with a nearby coronal hole, we extrapolated the global magnetic field using PFSS and looked for observational signatures of coronal holes in the AIA EUV images (Figure \ref{ch}). Figure \ref{ch}(b) shows a global PFSS extrapolation concentrating around the AR NOAA 12673 (the AR is indicated by a black arrow in Figure \ref{ch}(b)), where open and closed field lines are shown in grey and blue, respectively. From Figure \ref{ch}(b), we find that open field lines originated to the north of the flaring region close to the negative polarity, as well as to the west side trailing to the south and deflected toward the south-western direction. Here we recall that the CME associated with the M-class flares propagated along the same direction (see Figure \ref{cme}). Figures \ref{ch}(a) and (c) represent AIA 193 \AA\ images close to the peak of the first M1.4 flare on 2017 September 7 and one solar rotation later on 2017 October 4, respectively. There was no unambiguous observational signature of a coronal hole in the EUV images at the time of the events under study. However, one solar rotation later, a small but prominent coronal hole region was observed at the identified location of open field lines (indicated by the white arrow in Figure \ref{ch}(c)).

\subsection{Non-linear force-free field (NLFFF) extrapolation} \label{sec:nlfff}
\subsubsection{Optimization based NLFFF extrapolation technique}
To understand the coronal magnetic field configuration associated with the AR NOAA 12673, we applied an optimization technique \citep{Wheatland2000, Wiegelmann2004} to compute the NLFFF-equilibrium. Here we used an advanced version of this code, which takes care of measurement errors in the magnetogram \citep{Wiegelmann2010} and has been optimized for use with data from \textit{SDO}/HMI \citep{Wiegelmann2012}. In the optimization approach, $L$ is minimized \citep{Wiegelmann2010}, where
\begin{equation} \label{eq1}
L=\int_{V}^{} \left(\omega_f\frac{\mid(\nabla\times\vec{B})\times\vec{B}\mid^2}{B^2}+\omega_d\mid\nabla\cdot\vec{B}\mid^2\right)dv+\nu\int_{S}^{}(\vec{B}-\vec{B}_{obs})\cdot \mathbf{W} \cdot(\vec{B}-\vec{B}_{obs})d\vec{S}.
\end{equation}
Here, $\omega_f$, $\omega_d$, and $\nu$ are weighting functions while $\mathbf{W}$ is a diagonal error matrix with the elements $w_{los}$, $w_{trans}$, and $w_{trans}$; `$los$' and `$trans$' being the line-of-sight and transverse components, respectively. The NLFFF code used in this article, calculates $\omega_f$, $\omega_d$ (in the code, $\omega_f$ and $\omega_d$ are chosen to be identical i.e., $\omega_f=\omega_d$). and allows $\nu$, $w_{los}$, and $w_{trans}$ as free parameters (i.e., these parameters can be explicitly defined upon calling of the preprocessing/optimization). Since the photosphere is not force-free, the photospheric mangetograms used as the input boundary conditions, need to be pre-processed \citep{Wiegelmann2006}. In Equation \ref{eq1}, $\vec{B_{obs}}$ denotes the preprocessed magnetic field. For the purpose of preprocessing, a second functional $\mathscr{L}$ is defined as
\begin{equation} \label{eq2}
\mathscr{L}=\mu_1\mathscr{L}_1+\mu_2\mathscr{L}_2+\mu_3\mathscr{L}_3+\mu_4\mathscr{L}_4
\end{equation}
where
\begin{align*}
\mathscr{L}_1 &=\left(\sum B_xB_z\right)^2 + \left(\sum B_yB_z\right)^2 +\left(\sum B^2_z-B^2_x-B^2_y\right)^2 \\
\mathscr{L}_2 &=\left(\sum x(B^2_z-B^2_x-B^2_y)\right)^2+\left(\sum y(B^2_z-B^2_x-B^2_y)\right)^2+\left(\sum yB_xB_z-xB_yB_z\right)^2 \\
\mathscr{L}_3 &=\sum (B_x-B_{x,obs})^2+\sum (B_y-B_{y,obs})^2+\sum (B_z-B_{z,obs})^2 \\
\mathscr{L}_4 &=\sum \left((\Delta B_x)^2+(\Delta B_y)^2+(\Delta B_z)^2\right)
\end{align*}
Here, the summations are done over all the grid nodes of the bottom boundary. The values of $\mu_1$, $\mu_2$, $\mu_3$, and $\mu_4$ are free parameters and therefore, user defined.

In this article, for pre-processing the input photospheric magnetogram, we used the values of free parameters as follows:
\begin{equation} \label{eq3}
\nu=0.01; ~~~ w_{los}=1; ~~~ w_{trans}=\frac{B_{trans}}{max(B_{trans})};~~~\mu_1=\mu_2=1;~~~\mu_3=0.001;~~~\mu_4=0.01
\end{equation}
In Table \ref{pptable}, we compare the values of dimensionless flux, dimensionless force and dimensionless torque before and after pre-processing, which can be used to assess the degree of force-freeness of the input processed magnetograms. With the extrapolated magnetic field, the following parameters were calculated which can be considered as the quantification of force and divergence freeness of the extrapolated magnetic field:
\begin{equation} \label{eq4}
Fractional~flux~ratio\footnote{See \citet{DeRosa2015}.}~(<|f_i|>)=5.09\times10^{-4};~~\mid\vec{J}\times\vec{B}\mid=3.8\times10^{-3};~~~weighted~angle~between~ \vec{J}~ and~ \vec{B}=6.72^\circ
\end{equation}
\begin{deluxetable*}{P{4cm}P{3cm}P{3cm}}
\tablenum{2}
\tablecaption{Values of different parameters before and after pre-processing}
\label{pptable}
\centering
\tablehead{\colhead{Parameter} & \colhead{Before} & \colhead{After}}
\startdata
Flux balance & $-6.60\times$10$^{-3}$ & $-2.34\times$10$^{-6}$ \\
Dimensionless force & 0.29 & 4.18$\times$10$^{-4}$ \\
Dimensionless torque & 0.31 & 1.05$\times$10$^{-3}$ \\
\enddata
\end{deluxetable*}

In this context, it is noteworthy that at the time of extrapolation i.e., 2017 September 07 09:46 UT, the AR NOAA 12673 was centered at the heliographic position $\sim$S07W46\footnote{\url{https://www.lmsal.com/solarsoft/latest_events_archive/events_summary/2017/09/07/gev_20170907_0949/index.html}}. A detailed study by \citet{Gary1990} on the implications of the Sun's curvature on the magnetogram observations, suggest that full spherical geometry must be taken into account for off-center regions for angles $>$50$\degr$. During the extrapolations, we used `cylindrical equal area \citep[CEA;][]{Gary1990}' projected magnetograms which do not produce results as accurate as full spherical geometry does, but considering that we were interested in particularly the northeastern area of the AR which lies well within 50$\degr$, CEA projection can be accepted to be decently reliable.

\begin{figure}
   \centering
   \epsscale{1.2}
   \plotone{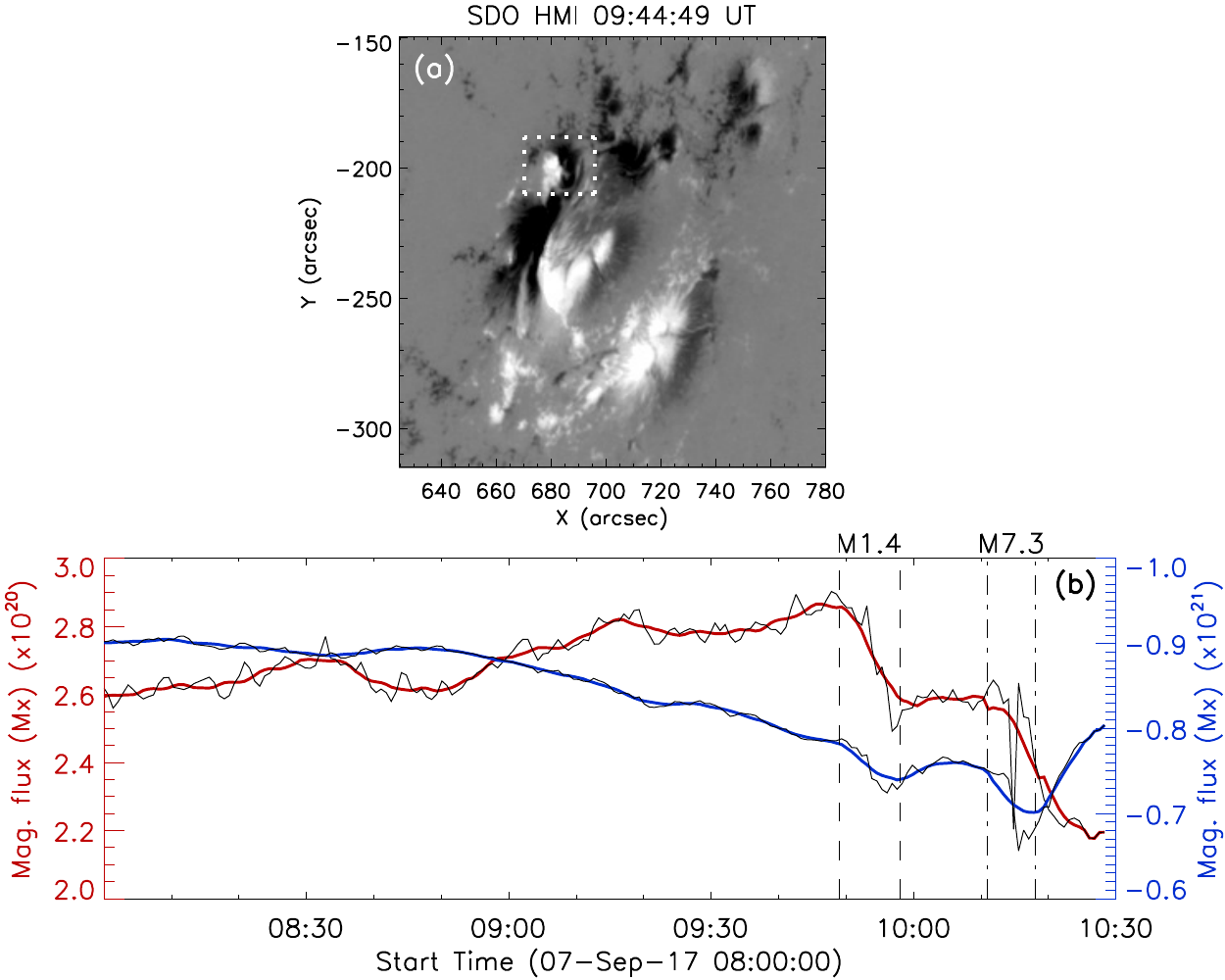}
   \caption{Panel (a): HMI Magnetogram of AR 12673 on 2017 September 7 09:45 UT. Time evolution of photospheric magnetic flux of the whole AR and inside the selected region within the dotted box in panel (a) are plotted in panel (b). The dashed lines in panel (b) mark the starting and ending time of the M1.4 flare as observed by GOES and the dashed-dotted lines in these mark the starting and ending time of the M7.3 flare as observed by GOES.}
   \label{hmi_lightcurve}
\end{figure}

\subsubsection{Extrapolation results}
In Figure \ref{nlfff}(a), we show a LOS magnetogram of the AR during the preflare phase. We specify a small part of the AR (shown inside the blue box in Figure \ref{nlfff}(a)) for plotting the NLFFF extrapolated field lines. The region inside the box represents a complex distribution of magnetic polarities in a largely bipolar configuration of major positive and negative fields in the western and eastern parts, respectively (the regions inside the blue box in Figure \ref{nlfff}(a)). In the northern part of the box, we find a small positive polarity region surrounded by negative polarity regions from three sides (the region inside the green box in Figure \ref{nlfff}(a)). In the north-western side of the AR, we find many disperse but strong negative polarity patches. For computation of the modeled magnetic field lines, we assume that the region showing flare associated brightenings (which also includes a filament) contains part of flux rope which underwent magnetic reconnection (i.e., relevant field lines).

NLFFF extrapolation results reveal the presence of two small MFRs along the PIL of the mini-sigmoid region (shown by blue and green lines in Figure \ref{nlfff}(b)). The two MFRs were intertwined with each other forming a ``double-decker flux rope system''. We plot only the intertwined MFRs in Figure \ref{nlfff}(c) for better visualization. Further, NLFFF extrapolation also suggests the presence of relatively large-scale closed magnetic field lines connecting the central positive and northern negative polarity regions (shown by the yellow lines in Figure \ref{nlfff}(b)).

\begin{figure}
   \centering
   \epsscale{0.5}
   \plotone{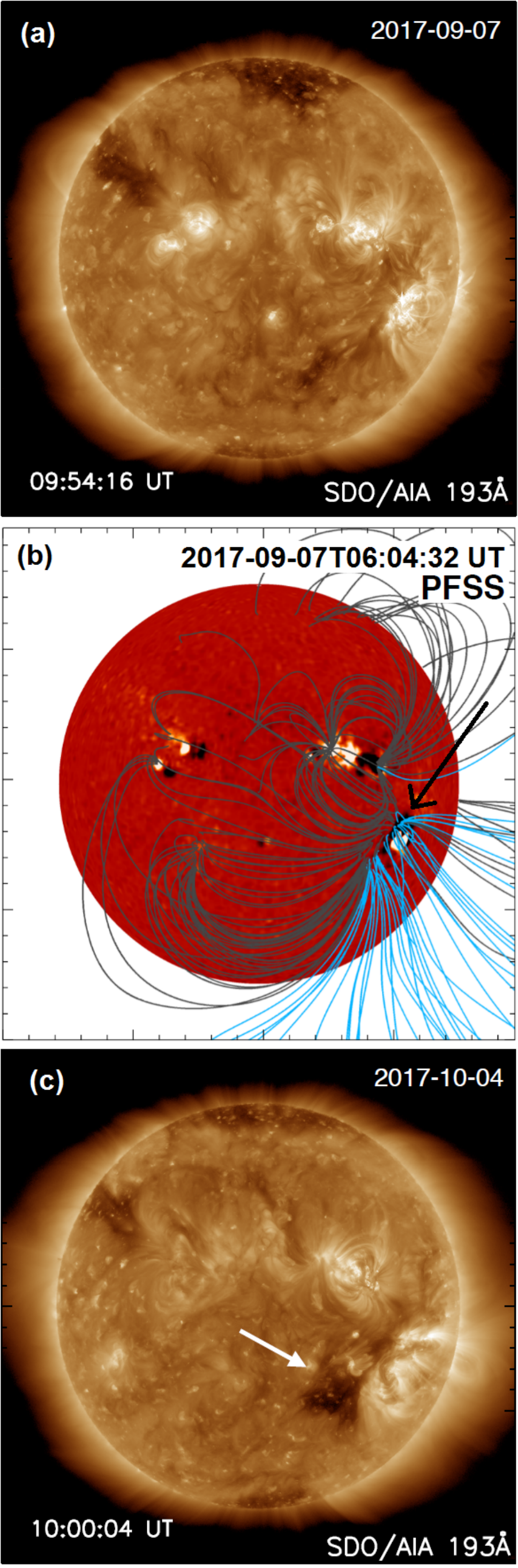}
   \caption{PFSS extrapolation of the global magnetic field (panel b) together with AIA 193 \AA\ images during the events under study (panel a) and one solar rotation later on October 4, 2017 (panel (c)). Grey and blue field lines indicate closed and open magnetic field lines, respectively. The open field lines originating to the north and west side of the flaring active region, are observed as the signature of a coronal hole, one solar rotation later (indicated by the white arrow).}
   \label{ch}
\end{figure}

\subsection{Distribution of magnetic decay index and twist number}
To explore, how the strength of the coronal magnetic field of AR 12673 varied with height, we calculated the magnetic decay index in the whole AR volume (i.e., 344$\times$224$\times$244 pixels; see Section \ref{data}). In Figures \ref{nlfff}(d), we show the variation of the magnetic decay index with height in a plane above the flux rope axis. For this purpose, we considered an approximate shape of the axis of the double-decker flux rope system which is shown by the red curves in Figures \ref{nlfff}(a) and (c). The approximated PIL was then projected onto the 2D lower boundary and the decay index was computed in a plane vertically above that approximate path. This process is similar to the technique undertaken by \citet{Liu2015}. We plot two contours on the vertical surface with levels $n$=1.0 and 1.5. The approximate height of the double-decker system is indicated by the yellow dashed-dotted line in Figure \ref{nlfff}(d). We find that a few segments of the double-decker flux ropes system were associated with a magnetic decay index as high as $\approx$1.0. In Figure \ref{nlfff}(e), we plot the variation of the decay index averaged over the path of the PIL as a function of height. Our results suggest that initially the decay index increased and reached a value of $\approx$0.9 within a height of $\approx$5 Mm. At larger heights, it experienced a sharp decrease up to the height of $\approx$9 Mm where the value of decay index (n) was $\approx$0.3. Above this height ($\approx$9 Mm), the average decay index was found to lie within the critical value (1.0--1.5) within the height range of 27--45 Mm.

In order to explore the possibility of kink instability as the triggering mechanism of the eruption of the flux rope, we calculated the twist number ($T_w$) in the flaring region, defined as \citep{Berger2006}
\begin{equation} \label{eq5}
T_w=\int_{L}\frac{(\nabla\times\vec{B})\cdot\vec{B}}{4\pi B^2}dl
\end{equation}
where $L$ is the length of the flux rope. Our calculations reveal that the double-decker system was associated with negative twist. The average value of $|T_w|$ associated with the double-decker system was found to be $\approx$1.0.

\section{Discussion} \label{discussion}
In this paper, we present a multiwavelengths analysis of two M-class flares from the AR NOAA 12673 on 2017 September 7 that resulted in the successive activation of a filament and subsequent narrow CME. As indicated in Figure \ref{goes}, both M-class flares were very impulsive with the respective impulsive phases lasting for $\approx$4 and $\approx$2 minutes only. EUV images of the AR revealed that the flaring activity occurred within a very localized region (indicated within the boxes in Figures \ref{intro}(c) and (e)). AIA 94 \AA\ images sampling hot coronal plasma, clearly revealed an inverted `S'-shaped structure lying in east-west orientation at the same location (Figure \ref{m1_4}(a)). Such coronal `S' (or, inverted `S') shaped structures are known as `coronal sigmoids' \citep[see,][]{Manoharan1996, Rust1996}. However, while usually coronal sigmoids are observed to have lengths of $\sim$100--300 Mm \citep[see, e.g.,][]{Tripathi2009, Joshi2017, Mitra2018, Mitra2020}, the sigmoidal structure reported in this article had a characteristic length of only $\sim$20 Mm ($\approx$30\arcsec). In view of the much smaller length-scales, we can be justifiably term it as `mini-sigmoid'. During different phases of the two M-class flares, we clearly observed the formation and activation of a small filament from the sigmoidal region and associated jet-like plasma ejection (Figures \ref{m1_4} and \ref{m7_3_2}).

\begin{figure}
   \centering
   \epsscale{0.98}
   \plotone{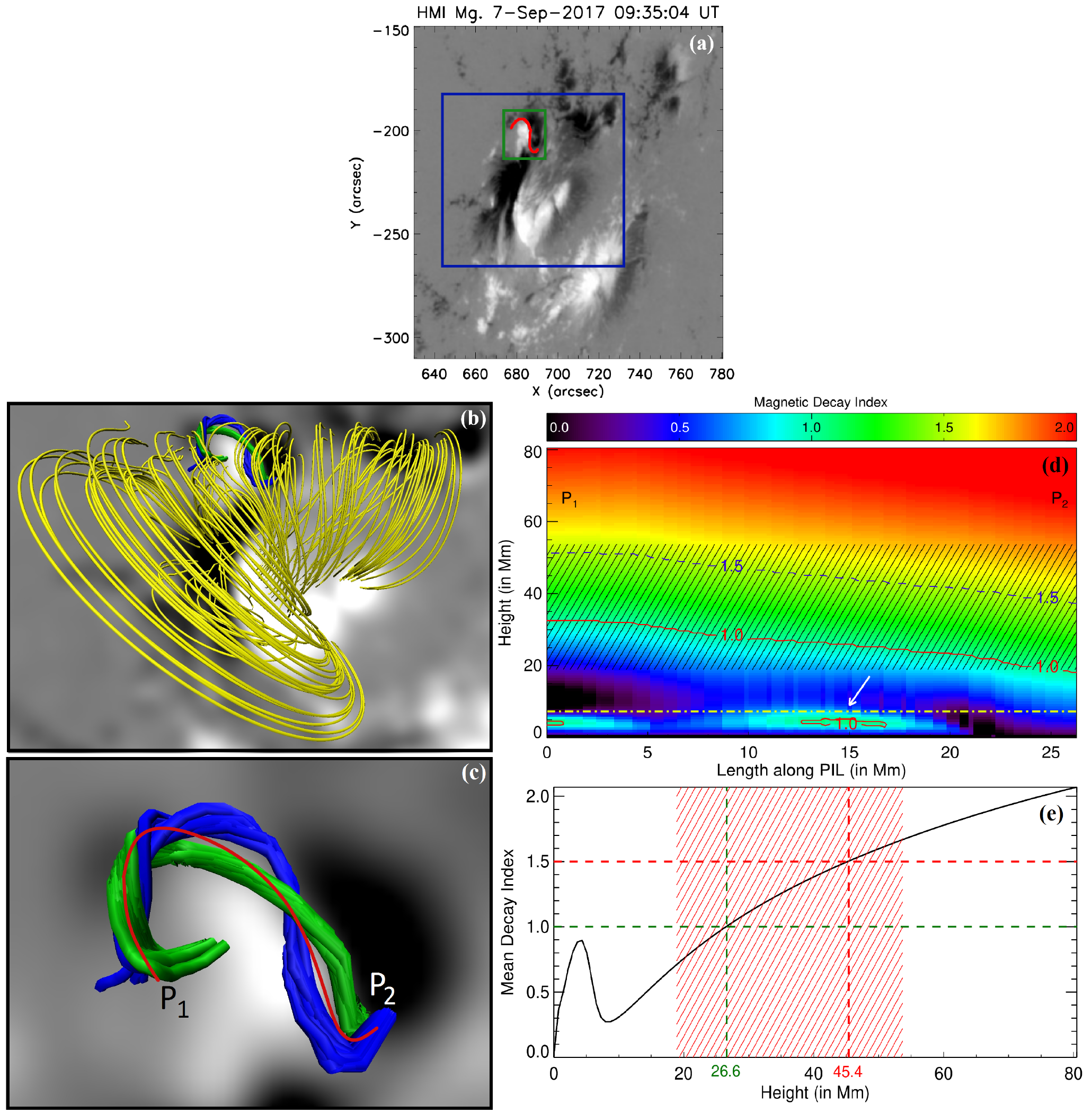}
   \caption{Panel (a): HMI LOS magnetogram showing the photospheric configuration of the active region NOAA 12673 prior to the flaring activity. Panel (b): NLFFF extrapolation results at 2017 September 07 09:45 UT showing coronal connectivities between different parts of the complex AR. Multiple flux ropes were identified in the extrapolation volume which are shown by blue and green lines. NLFFF extrapolated field lines are drawn over the photospheric region shown inside the blue box in panel (a). In panel (c) we only show the two flux ropes situated in the flaring region within the AR, constituting a double-decker flux rope system. The FOV of panel (c) is approximately indicated in panel (a) by the green box. Panel (d): distribution of magnetic decay index (n) above the PIL indicated by the red curves in panels (a) and (c). The yellow dashed-dotted line in panel (d) approximately indicate the height of the double-decker flux rope system. The blue dashed and red solid curves in panel (d) refer to contours of n=1.0 and 1.5, respectively. The white arrow in panel (d) indicate a region within the height of flux ropes, characterized by decay index n$\approx$1.0 which is higher than the surrounding. Panel (e): variation of mean decay index with height above the PIL. The green and red dashed lines in panel (e) mark the heights corresponding to n=1.0 and 1.5, respectively. The hatched region in panels (d) and (e) indicate the range of critical decay index height for torus instability as found by \citet{WangD2017}.}
   \label{nlfff}
\end{figure}

Sigmoids are associated with twisted or helical magnetic structures i.e., MFRs or filament channels \citep{Gibson2002}. MFRs are complex structures lying above PILs in the solar atmosphere where a set of magnetic field lines wrap around along its central axis more than once \citep{Gibson2006}. The results of the NLFFF extrapolation suggests the presence of two MFRs in the AR NOAA 12673 at the site of the M-class flares on 2017 September 7 (Figure \ref{nlfff}). Interestingly, the MFRs in the AR seem to wrap around each other forming a ``double-decker flux rope'' system. Double-decker flux rope system was first identified by \citet{Liu2012}. While studying an eruptive M1.0 class flare, they observed two vertically well separated filaments lying above a single PIL that remained stable for a few days before the upper branch erupted in association with an M-class flare. Based on their multipoint and multiwavelength analysis, they concluded that both filament branches emerged from beneath the photosphere with a vertical separation of $\approx$13 Mm between the two branches. Few hours before the eruptive flare, filament threads within the lower branch lifted up and merged with the upper branch, triggering its eruption. \citet{Cheng2014} reported another case of double-decker flux ropes associated with an X-class eruptive flare from the sigmoidal AR 11520. They found the primary MFR to be formed $\approx$40 hours before the flare via tether-cutting reconnection between two J-shaped arcades. The second flux rope became evident in the hot coronal channels just $\approx$2 hours before the flare and its eruption developed into a CME. The high temperature of the second flux rope led them to conclude  internal reconnection to be responsible for its formation. Notably, the two MFRs reported by \citet{Cheng2014} were intertwined with each other which is very similar to our case.

The sigmoidal structure reported in this study was sustained after the eruptive flares (Figures \ref{m7_3}(m)--(r)) which suggests that only one of the two MFRs from the double-decker system erupted during the M-class flares. The previously reported cases pertaining to the eruption of double-decker flux rope systems exhibited similar situations \citep{Cheng2014}. The specialty of our study lies in the spatial extent of the sigmoid and the MFRs. Whereas, signatures of the double-decker flux rope (or filament) system were clearly identified in optical and (E)UV images in the earlier studies, we were not able to resolve two distinct observable features of the MFRs associated with the ``double-decker'' configuration in direct images (Figure \ref{m1_4}(a)). We attribute this to the circumstances that the flaring region was very localized, making it hard to distinguish the two flux rope systems within the resolution of AIA besides possible projection effects (see Figures \ref{m1_4} and \ref{m7_3_2}). However, few of the brightenings observed in AIA 1600 \AA\ images of the preflare phase (see Figure \ref{intro}(f)) may possibly be the footpoints of the two MFRs of the double-decker configuration. It is worth mentioning that the mini-sigmoidal structure appeared in the AIA 94 \AA\ channel only $\approx$1.5 hours before the M1.4 flare which is much shorter time compared to the evolution of the sigmoidal structure associated with the double-decker flux rope system, reported previously by \citet{Cheng2014} which was $\approx$40 hours.

Both the M-class flares initiated with very localized brightenings at the sigmoid, immediately followed by a collimated plasma ejection (Figures \ref{m1_4} and \ref{m7_3_2}). During the evolution of the flares, we identified the appearance of a small filament while the localized brightening persisted beneath the filament (Figure \ref{p2}). The photospheric magnetogram and extrapolated coronal magnetic field configuration clearly revealed the source region of the eruption to be bipolar with high shearing (Figure \ref{nlfff}). Further, both the positive and the absolute negative flux decayed prior to the flares from the flaring region (Figure \ref{hmi_lightcurve}(b)). We interpret this evolution as observational signatures of photospheric flux cancellation at PIL which is recognized for its association with small-scale magnetic reconnections \citep{Van1989} leading to the formation of MFRs which was well elaborated in subsequent studies \citep[see e.g.,][]{Amari2010, Xue2017, Panesar2018, Mitra2020}. For both flares, localized (E)UV brightenings, underneath the apparent location of the filament body, were observed during their onset. We interpret these findings as evidences for the tether-cutting model of solar eruptions \citep{Moore1992, Moore2001}. Among the observable signatures of tether-cutting reconnection, compact EUV and HXR brightenings beneath an erupting MFR (or middle of the sigmoid) and collimated plasma outflows may be highlighted \citep{Raftery2010, Liu2013, Chen2014, Chen2016, Chen2018}. While investigating the onset processes of a solar eruption, \citet{Chen2018} observed clear signatures of flux cancellation from the flaring region, bidirectional jets, and change in the topology of the hot loops during the precursor phase. They argued that bidirectional jet-like flows occurred as a result of interaction of two coronal loop structures. This led them to conclude that the onset process of the eruption was tether-cutting reconnection. The events reported in this article evolved with unidirectional jets which differs from the bidirectional jets reported by \citet{Chen2018}. However, EUV images of the flaring region confirm that the location of the occurrences of the jets were closely associated with the initial brightenings beneath the filaments (Figures \ref{m1_4}(e), (m) and \ref{m7_3}(d)) which further supports the tether-cutting reconnection between the two MFRs in the double-decker flux rope system.

It is noteworthy that, both the M-class flares initiated with highly collimated, unidirectional plasma outflow (Figures \ref{m1_4}, \ref{m7_3_2}, \ref{m7_3}) for a relatively short duration ($\sim$5 min), a characteristic feature of coronal jets \citep[see,][]{Raouafi2016}. In the present observations, the jet structure observed during the first M-class flare (M1.4) was associated with a collimated and narrow spire while its base was rooted in the highly sheared double-decker flux rope system. Such configuration of the coronal jet is consistent with the standard jet scenario. On the other hand, the jet occurring during the second the M-class flare (M7.3) initiated like a standard jet but gradually moved toward the ``blowout" phase. Notably, we found a distinct time-gap between the ``standard'' and ``blowout'' phases of $\sim$2 min (Figures \ref{m7_3_2}, \ref{m7_3}) which is attributed to slow kinematic evolution of the filament constrained by the base arch \citep[see Section 3 in][]{Moore2010}. The intense and impulsive SXR flux, peaking at M7.3 level, essentially manifests reconnection and the heated field lines as the blowout eruption of the filament proceeds. The jet was eventually associated with the eruption of a filament that resulted in a narrow CME.

The plasma ejection along with the jet's spires during both M-class flares presented some atypical features, requiring further investigation. AIA images clearly displayed that a part of the ejected material, after moving toward east for a short period, sharply changed direction from east to south-west (Figures \ref{m1_4} and \ref{m7_3_2}). This anomalous dynamics of jet propagation was much more prominent during the second flare. Notably, PFSS extrapolation revealed a set of open field lines which originated near the flaring site and underwent bending toward the south-west direction (Figure \ref{ch}(b)). Such regions of open field lines have been identified as dark areas in the form of CHs observed in the EUV and SXR images of the Sun \citep{Cranmer2009}. CH regions are believed to have strong impact on CMEs when they interact with each other. A statistical study conducted by \citet{Gopalswamy2009} revealed that a significant number of CMEs are deflected away from their initial propagation direction by CHs. Later on, several other studies provided results in support of this finding \citep[see e.g.,][]{Mohamed2012, Kahler2012, Wood2012, Wang2014, Bilenko2017, Yang2018, Heinemann2019}. On 2017 September 7, we did not observe any CH near the AR NOAA 12673. However, a faint dark region, identified in AIA 193 \AA\ images, situated in the south-eastern direction of the AR (Figure \ref{ch}(a)) developed into a prominent CH after one solar rotation (Figure \ref{ch}(c)). From its association with the open field lines (Figure \ref{ch}(b)), we conclude that deflection of the ejected material during the M-class flares reported in this article, was caused by the open field lines originating from the emerging CH region at the trailing part of the AR.

The horizontal magnetic field above the AR experienced a rapid decay with height and the conditions of torus instability was reached within low atmospheric height (Figure \ref{nlfff}(d)). The condition of torus instability has been extensively used to explain the eruptive nature of ARs \citep[see e.g.,][]{LiuY2008, Aulanier2010, Demoulin2010, Thalmann2015, Thalmann2016, Zuccarello2017, Chandra2017b, LiuL2018, Sarkar2018}. From theoretical calculations \citet{Bateman1978} proposed that a toroidal current ring becomes unstable for expansion if the surrounding poloidal field decreases radially faster than a critical value (n=1.5). \citet{Kliem2006} generalized this idea and proposed that flux rope structures can attain eruptive motions under the condition of torus instability. Several observational and theoretical studies have revealed the critical value of magnetic decay index for torus instability to lie within the range [1.0, 1.5] \citep{Demoulin2010, Olmedo2010, LiuY2008, Zuccarello2014, Liu2015}. \citet{WangD2017} conducted a statistical survey in order to calculate from the observations the critical height ($h_{crit}$) of torus instability for 60 two-ribbon flares. Their study revealed that on average the critical height where the decay index reached a value of n=1.5 was $h_{crit}=36.3\pm17.4$ Mm above the PIL. During the preflare phase of the M-class flares reported in this article, the condition of torus instability ($n_{crit}=1.0-1.5$ was achieved at the heights of $h_{crit}\approx$27--45 Mm above the PIL, which are in basic agreement with the statistical results reported in \citet[][$h_{crit}=36.3\pm17.4$ Mm]{WangD2017} and \citet[][$h_{crit}=21\pm10$ Mm]{Baumgartner2018}.

We further explored the application of kink instability toward the triggering of the flux ropes by computing the twist numbers. Our analysis reveals the average twist associated with the flux ropes to be $|T_w|\approx$1.0. An extensive statistical work by \citet{Duan2019} concerning the torus and kink instabilities as the triggering mechanisms of flux ropes, revealed that the critical value for the onset of kink instability is given by a twist number $|T_w|\approx$2. In view of this result, we could not establish any conclusive evidence of kink instability as a possible triggering mechanism in our event.

In summary, this paper studies the initiation and evolution of two homologous M-class flares in the AR NOAA 12673, which produced the two largest flares the solar cycle 24. Both flares underwent very impulsive evolution. An interesting feature of these eruptive flares lies in their association with a mini-sigmoid region which suggests that sufficient energy storage within even smaller magnetic flux ropes can also lead to CME initiation, provided the overlying magnetic field configuration is favorable for its further expansion in the corona. Our analysis suggest the flaring region to be associated with a double-decker flux rope configuration which constitutes a more complex case of energy storage within a compact region. Both M-class flares initiated with jet-like plasma ejections. We find the activation and rise of a filament after the first flare which then erupted during the second flare. The eruption of the filament at the source region can be justifiably termed as ``anomalous'' as the initial jet-like eruption not only drastically changed its direction but also underwent a large angular expansion as the erupting plasma reached at successive higher coronal regions. From multiwavelength EUV imaging and PFSS magnetic field extrapolation, we showed the presence of large-scale open field structures, expanding toward the south-west of the AR. Our analysis reveals that the anomalous expansion of the CME at the source region is due to the deflection of erupting material by the large-scale open field lines. In view of the presence of low coronal double-decker flux ropes and compact EUV brightenings beneath a filament along with the magnetic flux cancellation observed at the PIL, our analysis supports the tether-cutting model of solar eruptions. Further, the distribution of the magnetic decay index above the PIL suggests a rapid decay of the field above the mini-sigmoid region implying favorable coronal conditions for the successful eruption of the flux rope, initially activated by the tether-cutting process.

\acknowledgments
We thank the \textit{SDO} and \textit{RHESSI} teams for their open data policy. \textit{SDO} is NASA's mission under the Living With a Star (LWS) program. \textit{RHESSI} was NASA's mission under the SMall EXplorer (SMEX) program. We also thank Julia K. Thalmann for help in the NLFFF extrapolation. This work is supported by the Indo-Austrian joint research project No. INT/AUSTRIA/BMWF/ P-05/2017 and OeAD project No. IN 03/2017. A.M.V and K.D also thank the Austrian Science Fund (FWF): P27292-N20. We also thank the anonymous referee for his/her constructive comments that helped us improve the overall quality of this article.


\begin{thebibliography}{}
\expandafter\ifx\csname natexlab\endcsname\relax\def\natexlab#1{#1}\fi

\bibitem[{Allen~Gary \& Hagyard(1990)}]{Gary1990}
Allen~Gary, G., \& Hagyard, M.~J. 1990, Solar Physics, 126, 21

\bibitem[{{Amari} {et~al.}(2010){Amari}, {Aly}, {Mikic}, \&
  {Linker}}]{Amari2010}
{Amari}, T., {Aly}, J.-J., {Mikic}, Z., \& {Linker}, J. 2010, \apjl, 717, L26

\bibitem[{{Antiochos} {et~al.}(1999){Antiochos}, {DeVore}, \&
  {Klimchuk}}]{Antiochos1999}
{Antiochos}, S.~K., {DeVore}, C.~R., \& {Klimchuk}, J.~A. 1999, \apj, 510, 485

\bibitem[{{Archontis} \& {Hood}(2013)}]{Archontis2013}
{Archontis}, V., \& {Hood}, A.~W. 2013, \apjl, 769, L21

\bibitem[{{Asai} {et~al.}(2008){Asai}, {Shibata}, {Hara}, \&
  {Nitta}}]{Asai2008}
{Asai}, A., {Shibata}, K., {Hara}, H., \& {Nitta}, N.~V. 2008, \apj, 673, 1188

\bibitem[{{Aulanier}(2014)}]{Aulanier2014}
{Aulanier}, G. 2014, in IAU Symposium, Vol. 300, Nature of Prominences and
  their Role in Space Weather, ed. B.~{Schmieder}, J.-M. {Malherbe}, \& S.~T.
  {Wu}, 184--196

\bibitem[{{Aulanier} {et~al.}(2013){Aulanier}, {D{\'e}moulin}, {Schrijver},
  {Janvier}, {Pariat}, \& {Schmieder}}]{Aulanier2013}
{Aulanier}, G., {D{\'e}moulin}, P., {Schrijver}, C.~J., {et~al.} 2013, \aap,
  549, A66

\bibitem[{{Aulanier} {et~al.}(2012){Aulanier}, {Janvier}, \&
  {Schmieder}}]{Aulanier2012}
{Aulanier}, G., {Janvier}, M., \& {Schmieder}, B. 2012, \aap, 543, A110

\bibitem[{{Aulanier} {et~al.}(2010){Aulanier}, {T{\"o}r{\"o}k}, {D{\'e}moulin},
  \& {DeLuca}}]{Aulanier2010}
{Aulanier}, G., {T{\"o}r{\"o}k}, T., {D{\'e}moulin}, P., \& {DeLuca}, E.~E.
  2010, \apj, 708, 314

\bibitem[{{Bateman}(1978)}]{Bateman1978}
{Bateman}, G. 1978, {MHD instabilities}

\bibitem[{{Baumgartner} {et~al.}(2018){Baumgartner}, {Thalmann}, \&
  {Veronig}}]{Baumgartner2018}
{Baumgartner}, C., {Thalmann}, J.~K., \& {Veronig}, A.~M. 2018, \apj, 853, 105

\bibitem[{{Benz}(2017)}]{Benz2017}
{Benz}, A.~O. 2017, Living Reviews in Solar Physics, 14, 2

\bibitem[{{Berger} \& {Prior}(2006)}]{Berger2006}
{Berger}, M.~A., \& {Prior}, C. 2006, Journal of Physics A Mathematical
  General, 39, 8321

\bibitem[{{Bhatnagar}(1996)}]{Bhatnagar1996}
{Bhatnagar}, A. 1996, \apss, 243, 105

\bibitem[{Bilenko(2017)}]{Bilenko2017}
Bilenko, I.~A. 2017, Geomagnetism and Aeronomy, 57, 952

\bibitem[{{Brueckner} \& {Bartoe}(1983)}]{Brueckner1983}
{Brueckner}, G.~E., \& {Bartoe}, J. D.~F. 1983, \apj, 272, 329

\bibitem[{Brueckner {et~al.}(1995)Brueckner, Howard, Koomen, Korendyke,
  Michels, Moses, Socker, Dere, Lamy, Llebaria, Bout, Schwenn, Simnett,
  Bedford, \& Eyles}]{Brueckner1995}
Brueckner, G.~E., Howard, R.~A., Koomen, M.~J., {et~al.} 1995, Solar Physics,
  162, 357

\bibitem[{{Carmichael}(1964)}]{Carmichael1964}
{Carmichael}, H. 1964, NASA Special Publication, 50, 451

\bibitem[{{Chandra} {et~al.}(2017){Chandra}, {Filippov}, {Joshi}, \&
  {Schmieder}}]{Chandra2017b}
{Chandra}, R., {Filippov}, B., {Joshi}, R., \& {Schmieder}, B. 2017, \solphys,
  292, 81

\bibitem[{{Chen} {et~al.}(2020){Chen}, {Yu}, {Reeves}, \& {Gary}}]{Chen2020}
{Chen}, B., {Yu}, S., {Reeves}, K.~K., \& {Gary}, D.~E. 2020, \apjl, 895, L50

\bibitem[{{Chen} {et~al.}(2018){Chen}, {Duan}, {Yang}, {Yang}, \&
  {Dai}}]{Chen2018}
{Chen}, H., {Duan}, Y., {Yang}, J., {Yang}, B., \& {Dai}, J. 2018, \apj, 869,
  78

\bibitem[{{Chen} {et~al.}(2014){Chen}, {Zhang}, {Cheng}, {Ma}, {Yang}, \&
  {Li}}]{Chen2014}
{Chen}, H., {Zhang}, J., {Cheng}, X., {et~al.} 2014, \apjl, 797, L15

\bibitem[{{Chen} {et~al.}(2016){Chen}, {Zhang}, {Li}, \& {Ma}}]{Chen2016}
{Chen}, H., {Zhang}, J., {Li}, L., \& {Ma}, S. 2016, \apjl, 818, L27

\bibitem[{{Cheng} {et~al.}(2014){Cheng}, {Ding}, {Zhang}, {Sun}, {Guo}, {Wang},
  {Kliem}, \& {Deng}}]{Cheng2014}
{Cheng}, X., {Ding}, M.~D., {Zhang}, J., {et~al.} 2014, \apj, 789, 93

\bibitem[{Clyne {et~al.}(2007)Clyne, Mininni, Norton, \& Rast}]{Clyne2007}
Clyne, J., Mininni, P., Norton, A., \& Rast, M. 2007, New Journal of Physics,
  9, 301

\bibitem[{Cranmer(2009)}]{Cranmer2009}
Cranmer, S.~R. 2009, Living Reviews in Solar Physics, 6, 3

\bibitem[{{Dalmasse} {et~al.}(2015){Dalmasse}, {Chandra}, {Schmieder}, \&
  {Aulanier}}]{Dalmasse2015}
{Dalmasse}, K., {Chandra}, R., {Schmieder}, B., \& {Aulanier}, G. 2015, \aap,
  574, A37

\bibitem[{{D{\'e}moulin} \& {Aulanier}(2010)}]{Demoulin2010}
{D{\'e}moulin}, P., \& {Aulanier}, G. 2010, \apj, 718, 1388

\bibitem[{{DeRosa} {et~al.}(2015){DeRosa}, {Wheatland}, {Leka}, {Barnes},
  {Amari}, {Canou}, {Gilchrist}, {Thalmann}, {Valori}, {Wiegelmann},
  {Schrijver}, {Malanushenko}, {Sun}, \& {R{\'e}gnier}}]{DeRosa2015}
{DeRosa}, M.~L., {Wheatland}, M.~S., {Leka}, K.~D., {et~al.} 2015, \apj, 811,
  107

\bibitem[{{Ding} {et~al.}(2002){Ding}, {Qiu}, \& {Wang}}]{Ding2002}
{Ding}, M.~D., {Qiu}, J., \& {Wang}, H. 2002, \apj, 576, L83

\bibitem[{{Domingo} {et~al.}(1995){Domingo}, {Fleck}, \&
  {Poland}}]{Domingo1995}
{Domingo}, V., {Fleck}, B., \& {Poland}, A.~I. 1995, \solphys, 162, 1

\bibitem[{{Duan} {et~al.}(2019){Duan}, {Jiang}, {He}, {Feng}, {Zou}, \&
  {Cui}}]{Duan2019}
{Duan}, A., {Jiang}, C., {He}, W., {et~al.} 2019, \apj, 884, 73

\bibitem[{{Fletcher} {et~al.}(2011){Fletcher}, {Dennis}, {Hudson}, {Krucker},
  {Phillips}, {Veronig}, {Battaglia}, {Bone}, {Caspi}, {Chen}, {Gallagher},
  {Grigis}, {Ji}, {Liu}, {Milligan}, \& {Temmer}}]{Fletcher2011}
{Fletcher}, L., {Dennis}, B.~R., {Hudson}, H.~S., {et~al.} 2011, \ssr, 159, 19

\bibitem[{{Gary} {et~al.}(2018){Gary}, {Chen}, {Dennis}, {Fleishman},
  {Hurford}, {Krucker}, {McTiernan}, {Nita}, {Shih}, {White}, \&
  {Yu}}]{Gary2018}
{Gary}, D.~E., {Chen}, B., {Dennis}, B.~R., {et~al.} 2018, \apj, 863, 83

\bibitem[{{Gibson} \& {Fan}(2006)}]{Gibson2006}
{Gibson}, S.~E., \& {Fan}, Y. 2006, Journal of Geophysical Research (Space
  Physics), 111, A12103

\bibitem[{{Gibson} {et~al.}(2002){Gibson}, {Fletcher}, {Del Zanna}, {Pike},
  {Mason}, {Mandrini}, {D{\'e}moulin}, {Gilbert}, {Burkepile}, {Holzer},
  {Alexander}, {Liu}, {Nitta}, {Qiu}, {Schmieder}, \& {Thompson}}]{Gibson2002}
{Gibson}, S.~E., {Fletcher}, L., {Del Zanna}, G., {et~al.} 2002, \apj, 574,
  1021

\bibitem[{{Gopalswamy} {et~al.}(2009){Gopalswamy}, {M{\"a}kel{\"a}}, {Xie},
  {Akiyama}, \& {Yashiro}}]{Gopalswamy2009}
{Gopalswamy}, N., {M{\"a}kel{\"a}}, P., {Xie}, H., {Akiyama}, S., \& {Yashiro},
  S. 2009, Journal of Geophysical Research (Space Physics), 114, A00A22

\bibitem[{{Gou} {et~al.}(2017){Gou}, {Veronig}, {Dickson}, {Hernand ez-Perez},
  \& {Liu}}]{Gou2017}
{Gou}, T., {Veronig}, A.~M., {Dickson}, E.~C., {Hernand ez-Perez}, A., \&
  {Liu}, R. 2017, \apjl, 845, L1

\bibitem[{{Green} {et~al.}(2018){Green}, {T{\"o}r{\"o}k}, {Vr{\v{s}}nak},
  {Manchester}, \& {Veronig}}]{Green2018}
{Green}, L.~M., {T{\"o}r{\"o}k}, T., {Vr{\v{s}}nak}, B., {Manchester}, W., \&
  {Veronig}, A. 2018, \ssr, 214, 46

\bibitem[{{Guo} {et~al.}(2018){Guo}, {Dumbovi{\'c}}, {Wimmer-Schweingruber},
  {Temmer}, {Lohf}, {Wang}, {Veronig}, {Hassler}, {Mays}, {Zeitlin},
  {Ehresmann}, {Witasse}, {Freiherr von Forstner}, {Heber}, {Holmstr{\"o}m}, \&
  {Posner}}]{Guo2018}
{Guo}, J., {Dumbovi{\'c}}, M., {Wimmer-Schweingruber}, R.~F., {et~al.} 2018,
  Space Weather, 16, 1156

\bibitem[{{Heinemann} {et~al.}(2019){Heinemann}, {Temmer}, {Farrugia},
  {Dissauer}, {Kay}, {Wiegelmann}, {Dumbovi{\'c}}, {Veronig}, {Podladchikova},
  {Hofmeister}, {Lugaz}, \& {Carcaboso}}]{Heinemann2019}
{Heinemann}, S.~G., {Temmer}, M., {Farrugia}, C.~J., {et~al.} 2019, \solphys,
  294, 121

\bibitem[{Hirayama(1974)}]{Hirayama1974}
Hirayama, T. 1974, Solar Physics, 34, 323

\bibitem[{{Hou} {et~al.}(2018){Hou}, {Zhang}, {Li}, {Yang}, \& {Li}}]{Hou2018}
{Hou}, Y.~J., {Zhang}, J., {Li}, T., {Yang}, S.~H., \& {Li}, X.~H. 2018, \aap,
  619, A100

\bibitem[{{Hurford} {et~al.}(2002){Hurford}, {Schmahl}, {Schwartz}, {Conway},
  {Aschwanden}, {Csillaghy}, {Dennis}, {Johns-Krull}, {Krucker}, {Lin},
  {McTiernan}, {Metcalf}, {Sato}, \& {Smith}}]{Hurford2002}
{Hurford}, G.~J., {Schmahl}, E.~J., {Schwartz}, R.~A., {et~al.} 2002, \solphys,
  210, 61

\bibitem[{{Janvier} {et~al.}(2014){Janvier}, {Aulanier}, {Bommier},
  {Schmieder}, {D{\'e}moulin}, \& {Pariat}}]{Janvier2014}
{Janvier}, M., {Aulanier}, G., {Bommier}, V., {et~al.} 2014, \apj, 788, 60

\bibitem[{{Janvier} {et~al.}(2013){Janvier}, {Aulanier}, {Pariat}, \&
  {D{\'e}moulin}}]{Janvier2013}
{Janvier}, M., {Aulanier}, G., {Pariat}, E., \& {D{\'e}moulin}, P. 2013, \aap,
  555, A77

\bibitem[{{Jiang} {et~al.}(2007){Jiang}, {Chen}, {Li}, {Shen}, \&
  {Yang}}]{Jiang2007}
{Jiang}, Y.~C., {Chen}, H.~D., {Li}, K.~J., {Shen}, Y.~D., \& {Yang}, L.~H.
  2007, \aap, 469, 331

\bibitem[{{Joshi} {et~al.}(2013){Joshi}, {Kushwaha}, {Cho}, \&
  {Veronig}}]{Joshi2013}
{Joshi}, B., {Kushwaha}, U., {Cho}, K.-S., \& {Veronig}, A.~M. 2013, \apj, 771,
  1

\bibitem[{{Joshi} {et~al.}(2016){Joshi}, {Kushwaha}, {Veronig}, \&
  {Cho}}]{Joshi2016}
{Joshi}, B., {Kushwaha}, U., {Veronig}, A.~M., \& {Cho}, K.-S. 2016, \apj, 832,
  130

\bibitem[{{Joshi} {et~al.}(2017{\natexlab{a}}){Joshi}, {Kushwaha}, {Veronig},
  {Dhara}, {Shanmugaraju}, \& {Moon}}]{Joshi2017}
{Joshi}, B., {Kushwaha}, U., {Veronig}, A.~M., {et~al.} 2017{\natexlab{a}},
  \apj, 834, 42

\bibitem[{{Joshi} {et~al.}(2017{\natexlab{b}}){Joshi}, {Thalmann}, {Mitra},
  {Chandra}, \& {Veronig}}]{Joshi2017b}
{Joshi}, B., {Thalmann}, J.~K., {Mitra}, P.~K., {Chandra}, R., \& {Veronig},
  A.~M. 2017{\natexlab{b}}, \apj, 851, 29

\bibitem[{Joshi {et~al.}(2012)Joshi, Veronig, Manoharan, \& Somov}]{Joshi2012}
Joshi, B., Veronig, A., Manoharan, P.~K., \& Somov, B.~V. 2012, in Multi-scale
  Dynamical Processes in Space and Astrophysical Plasmas, ed. M.~P. Leubner \&
  Z.~V{\"o}r{\"o}s (Berlin, Heidelberg: Springer Berlin Heidelberg), 29--41

\bibitem[{{Joshi} {et~al.}(2009){Joshi}, {Veronig}, {Cho}, {Bong}, {Somov},
  {Moon}, {Lee}, {Manoharan}, \& {Kim}}]{Joshi2009}
{Joshi}, B., {Veronig}, A., {Cho}, K.~S., {et~al.} 2009, \apj, 706, 1438

\bibitem[{{Kahler} {et~al.}(2012){Kahler}, {Akiyama}, \&
  {Gopalswamy}}]{Kahler2012}
{Kahler}, S.~W., {Akiyama}, S., \& {Gopalswamy}, N. 2012, \apj, 754, 100

\bibitem[{{Karpen} {et~al.}(2012){Karpen}, {Antiochos}, \&
  {DeVore}}]{Karpen2012}
{Karpen}, J.~T., {Antiochos}, S.~K., \& {DeVore}, C.~R. 2012, \apj, 760, 81

\bibitem[{{Kliem} \& {T{\"o}r{\"o}k}(2006)}]{Kliem2006}
{Kliem}, B., \& {T{\"o}r{\"o}k}, T. 2006, Physical Review Letters, 96, 255002

\bibitem[{{Kopp} \& {Pneuman}(1976)}]{Kopp1976}
{Kopp}, R.~A., \& {Pneuman}, G.~W. 1976, \solphys, 50, 85

\bibitem[{K\"{u}nzel(1960)}]{Kunzel1960}
K\"{u}nzel, H. 1960, Astronomische Nachrichten, 285, 271

\bibitem[{{Kushwaha} {et~al.}(2014){Kushwaha}, {Joshi}, {Cho}, {Veronig},
  {Tiwari}, \& {Mathew}}]{Kushwaha2014}
{Kushwaha}, U., {Joshi}, B., {Cho}, K.-S., {et~al.} 2014, \apj, 791, 23

\bibitem[{{Lemen} {et~al.}(2012){Lemen}, {Title}, {Akin}, {Boerner}, {Chou},
  {Drake}, {Duncan}, {Edwards}, {Friedlaender}, {Heyman}, {Hurlburt}, {Katz},
  {Kushner}, {Levay}, {Lindgren}, {Mathur}, {McFeaters}, {Mitchell}, {Rehse},
  {Schrijver}, {Springer}, {Stern}, {Tarbell}, {Wuelser}, {Wolfson}, {Yanari},
  {Bookbinder}, {Cheimets}, {Caldwell}, {Deluca}, {Gates}, {Golub}, {Park},
  {Podgorski}, {Bush}, {Scherrer}, {Gummin}, {Smith}, {Auker}, {Jerram},
  {Pool}, {Soufli}, {Windt}, {Beardsley}, {Clapp}, {Lang}, \&
  {Waltham}}]{Lemen2012}
{Lemen}, J.~R., {Title}, A.~M., {Akin}, D.~J., {et~al.} 2012, \solphys, 275, 17

\bibitem[{{Lin} {et~al.}(2002){Lin}, {Dennis}, {Hurford}, {Smith}, {Zehnder},
  {Harvey}, {Curtis}, {Pankow}, {Turin}, {Bester}, {Csillaghy}, {Lewis},
  {Madden}, {van Beek}, {Appleby}, {Raudorf}, {McTiernan}, {Ramaty}, {Schmahl},
  {Schwartz}, {Krucker}, {Abiad}, {Quinn}, {Berg}, {Hashii}, {Sterling},
  {Jackson}, {Pratt}, {Campbell}, {Malone}, {Landis}, {Barrington-Leigh},
  {Slassi-Sennou}, {Cork}, {Clark}, {Amato}, {Orwig}, {Boyle}, {Banks},
  {Shirey}, {Tolbert}, {Zarro}, {Snow}, {Thomsen}, {Henneck}, {McHedlishvili},
  {Ming}, {Fivian}, {Jordan}, {Wanner}, {Crubb}, {Preble}, {Matranga}, {Benz},
  {Hudson}, {Canfield}, {Holman}, {Crannell}, {Kosugi}, {Emslie}, {Vilmer},
  {Brown}, {Johns-Krull}, {Aschwanden}, {Metcalf}, \& {Conway}}]{Lin2002}
{Lin}, R.~P., {Dennis}, B.~R., {Hurford}, G.~J., {et~al.} 2002, \solphys, 210,
  3

\bibitem[{{Liu} {et~al.}(2013){Liu}, {Deng}, {Lee}, {Wiegelmann}, {Moore}, \&
  {Wang}}]{Liu2013}
{Liu}, C., {Deng}, N., {Lee}, J., {et~al.} 2013, \apjl, 778, L36

\bibitem[{Liu {et~al.}(2015)Liu, Deng, Liu, Lee, Pariat, Wiegelmann, Liu,
  Kleint, \& Wang}]{Liu2015}
Liu, C., Deng, N., Liu, R., {et~al.} 2015, \apjl, 812, L19

\bibitem[{{Liu} {et~al.}(2019){Liu}, {Cheng}, {Wang}, \& {Zhou}}]{Liu2019}
{Liu}, L., {Cheng}, X., {Wang}, Y., \& {Zhou}, Z. 2019, \apj, 884, 45

\bibitem[{{Liu} {et~al.}(2018{\natexlab{a}}){Liu}, {Cheng}, {Wang}, {Zhou},
  {Guo}, \& {Cui}}]{LiuL2018b}
{Liu}, L., {Cheng}, X., {Wang}, Y., {et~al.} 2018{\natexlab{a}}, \apj, 867, L5

\bibitem[{{Liu} {et~al.}(2018{\natexlab{b}}){Liu}, {Wang}, {Zhou}, {Dissauer},
  {Temmer}, \& {Cui}}]{LiuL2018}
{Liu}, L., {Wang}, Y., {Zhou}, Z., {et~al.} 2018{\natexlab{b}}, \apj, 858, 121

\bibitem[{Liu {et~al.}(2012)Liu, Kliem, Török, Liu, Titov, Lionello, Linker,
  \& Wang}]{Liu2012}
Liu, R., Kliem, B., Török, T., {et~al.} 2012, The Astrophysical Journal, 756,
  59

\bibitem[{{Liu} {et~al.}(2018{\natexlab{c}}){Liu}, {Jin}, {Downs}, {Ofman},
  {Cheung}, \& {Nitta}}]{LiuW2018}
{Liu}, W., {Jin}, M., {Downs}, C., {et~al.} 2018{\natexlab{c}}, \apj, 864, L24

\bibitem[{{Liu} {et~al.}(2008){Liu}, {Petrosian}, {Dennis}, \&
  {Jiang}}]{Liu2008}
{Liu}, W., {Petrosian}, V., {Dennis}, B.~R., \& {Jiang}, Y.~W. 2008, \apj, 676,
  704

\bibitem[{{Liu}(2008)}]{LiuY2008}
{Liu}, Y. 2008, \apjl, 679, L151

\bibitem[{{Manoharan} {et~al.}(1996){Manoharan}, {van Driel-Gesztelyi}, {Pick},
  \& {Demoulin}}]{Manoharan1996}
{Manoharan}, P.~K., {van Driel-Gesztelyi}, L., {Pick}, M., \& {Demoulin}, P.
  1996, \apjl, 468, L73

\bibitem[{{Maurya} \& {Ambastha}(2009)}]{Maurya2009}
{Maurya}, R.~A., \& {Ambastha}, A. 2009, \solphys, 258, 31

\bibitem[{{Maurya} {et~al.}(2012){Maurya}, {Vemareddy}, \&
  {Ambastha}}]{Maurya2012}
{Maurya}, R.~A., {Vemareddy}, P., \& {Ambastha}, A. 2012, \apj, 747, 134

\bibitem[{{Mitra} \& {Joshi}(2019)}]{Mitra2019}
{Mitra}, P.~K., \& {Joshi}, B. 2019, \apj, 884, 46

\bibitem[{{Mitra} {et~al.}(2020){Mitra}, {Joshi}, \& {Prasad}}]{Mitra2020}
{Mitra}, P.~K., {Joshi}, B., \& {Prasad}, A. 2020, \solphys, 295, 29

\bibitem[{Mitra {et~al.}(2018)Mitra, Joshi, Prasad, Veronig, \&
  Bhattacharyya}]{Mitra2018}
Mitra, P.~K., Joshi, B., Prasad, A., Veronig, A.~M., \& Bhattacharyya, R. 2018,
  ApJ, 869, 69

\bibitem[{{Mohamed} {et~al.}(2012){Mohamed}, {Gopalswamy}, {Yashiro},
  {Akiyama}, {M{\"a}kel{\"a}}, {Xie}, \& {Jung}}]{Mohamed2012}
{Mohamed}, A.~A., {Gopalswamy}, N., {Yashiro}, S., {et~al.} 2012, Journal of
  Geophysical Research (Space Physics), 117, A01103

\bibitem[{{Moore} {et~al.}(2010){Moore}, {Cirtain}, {Sterling}, \&
  {Falconer}}]{Moore2010}
{Moore}, R.~L., {Cirtain}, J.~W., {Sterling}, A.~C., \& {Falconer}, D.~A. 2010,
  \apj, 720, 757

\bibitem[{{Moore} \& {Roumeliotis}(1992)}]{Moore1992}
{Moore}, R.~L., \& {Roumeliotis}, G. 1992, in Lecture Notes in Physics, Berlin
  Springer Verlag, Vol. 399, IAU Colloq. 133: Eruptive Solar Flares, ed.
  Z.~{Svestka}, B.~V. {Jackson}, \& M.~E. {Machado}, 69

\bibitem[{{Moore} {et~al.}(2001){Moore}, {Sterling}, {Hudson}, \&
  {Lemen}}]{Moore2001}
{Moore}, R.~L., {Sterling}, A.~C., {Hudson}, H.~S., \& {Lemen}, J.~R. 2001,
  \apj, 552, 833

\bibitem[{{Moore} {et~al.}(1977){Moore}, {Tang}, {Bohlin}, \&
  {Golub}}]{Moore1977}
{Moore}, R.~L., {Tang}, F., {Bohlin}, J.~D., \& {Golub}, L. 1977, \apj, 218,
  286

\bibitem[{{Moraitis} {et~al.}(2019){Moraitis}, {Sun}, {Pariat}, \&
  {Linan}}]{Moraitis2019}
{Moraitis}, K., {Sun}, X., {Pariat}, {\'E}., \& {Linan}, L. 2019, \aap, 628,
  A50

\bibitem[{{Olmedo} \& {Zhang}(2010)}]{Olmedo2010}
{Olmedo}, O., \& {Zhang}, J. 2010, \apj, 718, 433

\bibitem[{{Panesar} {et~al.}(2018){Panesar}, {Sterling}, \&
  {Moore}}]{Panesar2018}
{Panesar}, N.~K., {Sterling}, A.~C., \& {Moore}, R.~L. 2018, \apj, 853, 189

\bibitem[{Pesnell {et~al.}(2012)Pesnell, Thompson, \& Chamberlin}]{Pesnell2012}
Pesnell, W.~D., Thompson, B.~J., \& Chamberlin, P.~C. 2012, Solar Physics, 275,
  3

\bibitem[{Priest \& Forbes(2000)}]{Priest2000}
Priest, E., \& Forbes, T. 2000, Magnetic Reconnection: MHD Theory and
  Applications (Cambridge University Press)

\bibitem[{{Priest} \& {Forbes}(2002)}]{Priest2002}
{Priest}, E.~R., \& {Forbes}, T.~G. 2002, \aapr, 10, 313

\bibitem[{{Raftery} {et~al.}(2010){Raftery}, {Gallagher}, {McAteer}, {Lin}, \&
  {Delahunt}}]{Raftery2010}
{Raftery}, C.~L., {Gallagher}, P.~T., {McAteer}, R.~T.~J., {Lin}, C.-H., \&
  {Delahunt}, G. 2010, \apj, 721, 1579

\bibitem[{{Raouafi} {et~al.}(2016){Raouafi}, {Patsourakos}, {Pariat}, {Young},
  {Sterling}, {Savcheva}, {Shimojo}, {Moreno-Insertis}, {DeVore}, {Archontis},
  {T{\"o}r{\"o}k}, {Mason}, {Curdt}, {Meyer}, {Dalmasse}, \&
  {Matsui}}]{Raouafi2016}
{Raouafi}, N.~E., {Patsourakos}, S., {Pariat}, E., {et~al.} 2016, \ssr, 201, 1

\bibitem[{Riley {et~al.}(2008)Riley, Lionello, Miki{\'{c}}, \&
  Linker}]{Riley2008}
Riley, P., Lionello, R., Miki{\'{c}}, Z., \& Linker, J. 2008, The Astrophysical
  Journal, 672, 1221

\bibitem[{{Romano} {et~al.}(2018){Romano}, {Elmhamdi}, {Falco}, {Costa},
  {Kordi}, {Al-Trabulsy}, \& {Al-Shammari}}]{Romano2018}
{Romano}, P., {Elmhamdi}, A., {Falco}, M., {et~al.} 2018, \apj, 852, L10

\bibitem[{{Romano} {et~al.}(2019){Romano}, {Elmhamdi}, \& {Kordi}}]{Romano2019}
{Romano}, P., {Elmhamdi}, A., \& {Kordi}, A.~S. 2019, \solphys, 294, 4

\bibitem[{{Rust} \& {Kumar}(1996)}]{Rust1996}
{Rust}, D.~M., \& {Kumar}, A. 1996, \apjl, 464, L199

\bibitem[{Sahu {et~al.}(2020)Sahu, Joshi, Mitra, Veronig, \&
  Yurchyshyn}]{Sahu2020}
Sahu, S., Joshi, B., Mitra, P.~K., Veronig, A.~M., \& Yurchyshyn, V. 2020, The
  Astrophysical Journal, 897, 157

\bibitem[{{Sarkar} \& {Srivastava}(2018)}]{Sarkar2018}
{Sarkar}, R., \& {Srivastava}, N. 2018, \solphys, 293, 16

\bibitem[{{Schmieder} {et~al.}(1995){Schmieder}, {Shibata}, {van
  Driel-Gesztelyi}, \& {Freeland}}]{Schmieder1995}
{Schmieder}, B., {Shibata}, K., {van Driel-Gesztelyi}, L., \& {Freeland}, S.
  1995, \solphys, 156, 245

\bibitem[{{Schou} {et~al.}(2012){Schou}, {Scherrer}, {Bush}, {Wachter},
  {Couvidat}, {Rabello-Soares}, {Bogart}, {Hoeksema}, {Liu}, {Duvall}, {Akin},
  {Allard}, {Miles}, {Rairden}, {Shine}, {Tarbell}, {Title}, {Wolfson},
  {Elmore}, {Norton}, \& {Tomczyk}}]{Schou2012}
{Schou}, J., {Scherrer}, P.~H., {Bush}, R.~I., {et~al.} 2012, \solphys, 275,
  229

\bibitem[{{Seaton} \& {Darnel}(2018)}]{Seaton2018}
{Seaton}, D.~B., \& {Darnel}, J.~M. 2018, \apj, 852, L9

\bibitem[{{Seehafer}(1978)}]{Seehafer1978}
{Seehafer}, N. 1978, \solphys, 58, 215

\bibitem[{{Shibata} \& {Magara}(2011)}]{Shibata2011}
{Shibata}, K., \& {Magara}, T. 2011, Living Reviews in Solar Physics, 8, 6

\bibitem[{{Shibata} {et~al.}(1996){Shibata}, {Yokoyama}, \&
  {Shimojo}}]{Shibata1996}
{Shibata}, K., {Yokoyama}, T., \& {Shimojo}, M. 1996, Journal of Geomagnetism
  and Geoelectricity, 48, 19

\bibitem[{{Shibata} {et~al.}(1992){Shibata}, {Ishido}, {Acton}, {Strong},
  {Hirayama}, {Uchida}, {McAllister}, {Matsumoto}, {Tsuneta}, {Shimizu},
  {Hara}, {Sakurai}, {Ichimoto}, {Nishino}, \& {Ogawara}}]{Shibata1992}
{Shibata}, K., {Ishido}, Y., {Acton}, L.~W., {et~al.} 1992, \pasj, 44, L173

\bibitem[{{Shimojo} {et~al.}(1996){Shimojo}, {Hashimoto}, {Shibata},
  {Hirayama}, {Hudson}, \& {Acton}}]{Shimojo1996}
{Shimojo}, M., {Hashimoto}, S., {Shibata}, K., {et~al.} 1996, \pasj, 48, 123

\bibitem[{{Shimojo} {et~al.}(1998){Shimojo}, {Shibata}, \&
  {Harvey}}]{Shimojo1998}
{Shimojo}, M., {Shibata}, K., \& {Harvey}, K.~L. 1998, \solphys, 178, 379

\bibitem[{{Sturrock}(1966)}]{Sturrock1966}
{Sturrock}, P.~A. 1966, \nat, 211, 695

\bibitem[{{Thalmann} {et~al.}(2015){Thalmann}, {Su}, {Temmer}, \&
  {Veronig}}]{Thalmann2015}
{Thalmann}, J.~K., {Su}, Y., {Temmer}, M., \& {Veronig}, A.~M. 2015, \apjl,
  801, L23

\bibitem[{{Thalmann} {et~al.}(2016){Thalmann}, {Su}, {Temmer}, \&
  {Veronig}}]{Thalmann2016}
{Thalmann}, J.~K., {Su}, Y., {Temmer}, M., \& {Veronig}, A.~M. 2016, in IAU
  Symposium, Vol. 320, Solar and Stellar Flares and their Effects on Planets,
  ed. A.~G. {Kosovichev}, S.~L. {Hawley}, \& P.~{Heinzel}, 60--63

\bibitem[{{Titov} {et~al.}(2002){Titov}, {Hornig}, \&
  {D{\'e}moulin}}]{Titov2002}
{Titov}, V.~S., {Hornig}, G., \& {D{\'e}moulin}, P. 2002, Journal of
  Geophysical Research (Space Physics), 107, 1164

\bibitem[{{T{\"o}r{\"o}k} {et~al.}(2004){T{\"o}r{\"o}k}, {Kliem}, \&
  {Titov}}]{Torok2004}
{T{\"o}r{\"o}k}, T., {Kliem}, B., \& {Titov}, V.~S. 2004, \aap, 413, L27

\bibitem[{{Tripathi} {et~al.}(2009){Tripathi}, {Kliem}, {Mason}, {Young}, \&
  {Green}}]{Tripathi2009}
{Tripathi}, D., {Kliem}, B., {Mason}, H.~E., {Young}, P.~R., \& {Green}, L.~M.
  2009, \apjl, 698, L27

\bibitem[{{Tsuneta} {et~al.}(1991){Tsuneta}, {Acton}, {Bruner}, {Lemen},
  {Brown}, {Caravalho}, {Catura}, {Freeland}, {Jurcevich}, {Morrison},
  {Ogawara}, {Hirayama}, \& {Owens}}]{Tsuneta1991}
{Tsuneta}, S., {Acton}, L., {Bruner}, M., {et~al.} 1991, \solphys, 136, 37

\bibitem[{{van Ballegooijen} \& {Martens}(1989)}]{Van1989}
{van Ballegooijen}, A.~A., \& {Martens}, P.~C.~H. 1989, \apj, 343, 971

\bibitem[{{Verma}(2018)}]{Verma2018}
{Verma}, M. 2018, \aap, 612, A101

\bibitem[{{Veronig} {et~al.}(2002){Veronig}, {Vr{\v{s}}nak}, {Temmer}, \&
  {Hanslmeier}}]{Veronig2002}
{Veronig}, A., {Vr{\v{s}}nak}, B., {Temmer}, M., \& {Hanslmeier}, A. 2002,
  \solphys, 208, 297

\bibitem[{{Veronig} {et~al.}(2006){Veronig}, {Karlick{\'y}}, {Vr{\v s}nak},
  {Temmer}, {Magdaleni{\'c}}, {Dennis}, {Otruba}, \& {P{\"o}tzi}}]{Veronig2006}
{Veronig}, A.~M., {Karlick{\'y}}, M., {Vr{\v s}nak}, B., {et~al.} 2006, \aap,
  446, 675

\bibitem[{{Veronig} {et~al.}(2018){Veronig}, {Podladchikova}, {Dissauer},
  {Temmer}, {Seaton}, {Long}, {Guo}, {Vr{\v{s}}nak}, {Harra}, \&
  {Kliem}}]{Veronig2018}
{Veronig}, A.~M., {Podladchikova}, T., {Dissauer}, K., {et~al.} 2018, \apj,
  868, 107

\bibitem[{{Vr{\v{s}}nak}(2016)}]{Vrsnak2016}
{Vr{\v{s}}nak}, B. 2016, Astronomische Nachrichten, 337, 1002

\bibitem[{{Vr{\v{s}}nak} {et~al.}(2004){Vr{\v{s}}nak}, {Mari{\v{c}}i{\'c}},
  {Stanger}, \& {Veronig}}]{Vrsnak2004}
{Vr{\v{s}}nak}, B., {Mari{\v{c}}i{\'c}}, D., {Stanger}, A.~L., \& {Veronig}, A.
  2004, \solphys, 225, 355

\bibitem[{{Wang} {et~al.}(2017){Wang}, {Liu}, {Wang}, {Liu}, {Chen}, {Liu},
  {Zhou}, \& {Zhang}}]{WangD2017}
{Wang}, D., {Liu}, R., {Wang}, Y., {et~al.} 2017, \apjl, 843, L9

\bibitem[{{Wang} {et~al.}(2014){Wang}, {Wang}, {Shen}, {Shen}, \&
  {Lugaz}}]{Wang2014}
{Wang}, Y., {Wang}, B., {Shen}, C., {Shen}, F., \& {Lugaz}, N. 2014, Journal of
  Geophysical Research (Space Physics), 119, 5117

\bibitem[{{Wang} \& {Sheeley}(1992)}]{Wang1992}
{Wang}, Y.-M., \& {Sheeley}, Jr., N.~R. 1992, \apj, 392, 310

\bibitem[{{Wheatland} {et~al.}(2000){Wheatland}, {Sturrock}, \&
  {Roumeliotis}}]{Wheatland2000}
{Wheatland}, M.~S., {Sturrock}, P.~A., \& {Roumeliotis}, G. 2000, \apj, 540,
  1150

\bibitem[{{Wiegelmann}(2004)}]{Wiegelmann2004}
{Wiegelmann}, T. 2004, \solphys, 219, 87

\bibitem[{{Wiegelmann} \& {Inhester}(2010)}]{Wiegelmann2010}
{Wiegelmann}, T., \& {Inhester}, B. 2010, \aap, 516, A107

\bibitem[{{Wiegelmann} {et~al.}(2006){Wiegelmann}, {Inhester}, \&
  {Sakurai}}]{Wiegelmann2006}
{Wiegelmann}, T., {Inhester}, B., \& {Sakurai}, T. 2006, \solphys, 233, 215

\bibitem[{{Wiegelmann} {et~al.}(2012){Wiegelmann}, {Thalmann}, {Inhester},
  {Tadesse}, {Sun}, \& {Hoeksema}}]{Wiegelmann2012}
{Wiegelmann}, T., {Thalmann}, J.~K., {Inhester}, B., {et~al.} 2012, \solphys,
  281, 37

\bibitem[{{Wood} {et~al.}(2012){Wood}, {Wu}, {Rouillard}, {Howard}, \&
  {Socker}}]{Wood2012}
{Wood}, B.~E., {Wu}, C.~C., {Rouillard}, A.~P., {Howard}, R.~A., \& {Socker},
  D.~G. 2012, \apj, 755, 43

\bibitem[{{Xue} {et~al.}(2017){Xue}, {Yan}, {Yang}, {Wang}, \&
  {Zhao}}]{Xue2017}
{Xue}, Z., {Yan}, X., {Yang}, L., {Wang}, J., \& {Zhao}, L. 2017, \apjl, 840,
  L23

\bibitem[{{Yang} {et~al.}(2018){Yang}, {Dai}, {Chen}, {Li}, \&
  {Jiang}}]{Yang2018}
{Yang}, J., {Dai}, J., {Chen}, H., {Li}, H., \& {Jiang}, Y. 2018, \apj, 862, 86

\bibitem[{{Yang} {et~al.}(2017){Yang}, {Zhang}, {Zhu}, \& {Song}}]{Yang2017}
{Yang}, S., {Zhang}, J., {Zhu}, X., \& {Song}, Q. 2017, \apj, 849, L21

\bibitem[{{Yashiro} {et~al.}(2004){Yashiro}, {Gopalswamy}, {Michalek}, {St.
  Cyr}, {Plunkett}, {Rich}, \& {Howard}}]{Yashiro2004}
{Yashiro}, S., {Gopalswamy}, N., {Michalek}, G., {et~al.} 2004, Journal of
  Geophysical Research (Space Physics), 109, A07105

\bibitem[{Zhao {et~al.}(2009)Zhao, Wang, Matthews, Ding, Zhao, \&
  Jin}]{Zhao2009}
Zhao, M., Wang, J.-X., Matthews, S., {et~al.} 2009, Research in Astronomy and
  Astrophysics, 9, 812

\bibitem[{{Zuccarello} {et~al.}(2017){Zuccarello}, {Chandra}, {Schmieder},
  {Aulanier}, \& {Joshi}}]{Zuccarello2017}
{Zuccarello}, F.~P., {Chandra}, R., {Schmieder}, B., {Aulanier}, G., \&
  {Joshi}, R. 2017, \aap, 601, A26

\bibitem[{Zuccarello {et~al.}(2014)Zuccarello, Seaton, Filippov, Mierla,
  Poedts, Rachmeler, Romano, \& Zuccarello}]{Zuccarello2014}
Zuccarello, F.~P., Seaton, D.~B., Filippov, B., {et~al.} 2014, The
  Astrophysical Journal, 795, 175

\end{thebibliography}
\end{document}